\documentclass{raa_twocolumn}

\usepackage{graphicx,times}             
\usepackage{natbib}
\usepackage{float}
\usepackage{soul}
\usepackage{amssymb,amsmath}
\bibpunct{(}{)}{;}{a}{}{,}

\usepackage[utf8]{inputenc}
\usepackage{amsmath}
\usepackage{amssymb}
\usepackage{hyperref}
\usepackage{xcolor}
\usepackage{listings}
\lstdefinestyle{terminal}{
  basicstyle=\ttfamily\small,
  backgroundcolor=\color{gray!10},
  frame=single,
  columns=fullflexible,
  breaklines=true,
  showstringspaces=false
}
\usepackage{natbib}
\usepackage{float}
\usepackage{soul}

\newcommand{\mysec}{Sec.~}
\newcommand{\myfig}{Fig.~}

\newcommand{\myeq}{Eq.~}
\newcommand\pyIGIMF{{\tt pyIGIMF}}

\newcommand{\diff}[0]{\mathrm{d}}

\defcitealias{Kroupa+2023}{Kroupa, Gjergo, \emph{et al.}, (2023)}
\defcitealias{kroupa+2024}{Kroupa, Gjergo, Jerabkova \& Yan (2024)}

\graphicspath{{./}{figures/}}

\begin{document}

\title{ Massive Star Formation at Supersolar Metallicities: Constraints on the Initial Mass Function}

   \volnopage{Vol.0 (20xx) No.0, 000--000}      
   \setcounter{page}{1}          

   \author{Eda Gjergo$^{\star}$ 
      \inst{1,2}
   \and Zhiyu Zhang$^{\dagger}$ 
      \inst{1,2}
   \and Pavel Kroupa
      \inst{3,4}
   \and Aleksei Sorokin         
      \inst{6}
   \and Zhiqiang Yan            
      \inst{1,2}
   \and Ziyi Guo            
      \inst{1,2}
   \and Tereza Jerabkova        
      \inst{5}
   \and Akram Hasani Zoonozi     
      \inst{3,7}
   \and Hosein Haghi             
      \inst{3,7,8}
   }

\institute{School of Astronomy and Space Science, Nanjing University, Nanjing 210093, China; {\\ \it $^{\star}$eda.gjergo@gmail.com, $^{\dagger}$zzhang@nju.edu.cn}   \\
        \and 
           Key Laboratory of Modern Astronomy and Astrophysics (Nanjing University),\\ Ministry of Education, Nanjing 210093, People's Republic of China.
        \and 
           Helmholtz-Institut f\"ur Strahlen und Kernphysik, Universit\"{a}t Bonn,\\ Nussallee 14-16, Bonn, North Rhine-Westphalia, D-53115, Germany.
        \and 
           Charles University in Prague, Faculty of Mathematics and Physics, Astronomical Institute,\\ V Hole\v{s}ovi\v{c}kách 2, CZ-180 00 Praha 8, Czech Republic
        \and 
           Department of Theoretical Physics and Astrophysics, Faculty of Science, Masaryk University, \\Kotl\'{a}\v{r}sk\'{a} 2, Brno 611 37, Czech Republic
        \and 
           Department of Applied Mathematics, Illinois Institute of Technology, 10 W. $35^{\text{\small nd}}$ St., Chicago IL 60616, USA
         \and 
           Department of Physics, Institute for Advanced Studies in Basic Sciences (IASBS), 11365-9161 Zanjan, Iran 
         \and 
           School of Astronomy, Institute for Research in Fundamental Sciences (IPM), 19395-5531 Tehran, Iran 
 \vs\no
   {\\ \small Received 20xx month day; accepted 20xx month day}
           }

\abstract{ 
Metals enhance the cooling efficiency of molecular clouds, promoting
fragmentation. Consequently, increasing the metallicity may boost the formation
of low-mass stars. Within the integrated galaxy initial mass function (IGIMF)
theory, this effect is empirically captured by a linear relation between the
slope of the low-mass stellar IMF, $\alpha_1$, and the metal mass fraction, $Z$.
This linear $\alpha_1$-$Z$ relation has been calibrated up to $\approx 2 \,
Z_{\odot}$, though higher metallicity environments are known to exist. We show
that if the linear $\alpha_1$-$Z$ relation extends to higher metallicities ($[Z]
\gtrsim 0.5$),  massive star formation is suppressed entirely. Alternatively,
fragmentation efficiency may saturate beyond some metallicity threshold if
gravitational collapse cascades rapidly enough. To model this behavior, we
propose a logistic function describing the transition from metallicity-sensitive
to metallicity-insensitive fragmentation regimes. We provide a user-friendly
public code, \pyIGIMF, which enables the instantaneous computation of the IGIMF
theory with the logistic $\alpha_1$-$Z$ relation. 
}


\authorrunning{E. Gjergo et al. }            
\titlerunning{Constraints on the IMF at supersolar Metallicity}  

\maketitle

\section{Introduction} \label{sec:intro}

Stars form in the coldest and densest clumps of molecular clouds (hereafter \emph{molecular cloud clumps}, \citealt{McKeeOstriker2007,BerginTafalla2007}). Each
clump produces a complete ensemble of stars, an \emph{embedded cluster},
constituting a single stellar population \citep{LadaLada2003,PortegiesZwart2010}.
Clumps that produce embedded clusters are sub-pc structures sufficiently massive to
undergo collapse under self-gravitation \citep[][their Sec.~3.1]{kroupa+2024}.
With their short lifetimes (usually $\lesssim 1$~Myr), embedded clusters are the
cradles of individual star-forming events \citep[for reviews, see
][]{kroupa+2013, Hopkins2018, kroupa+2024}.

The number distribution of stellar masses, $\xi_{\star}(m) = \diff N / \diff m$,
for stars formed in a
single star-forming event is known as the \emph{stellar~initial~mass~function}
(sIMF). Over its lifetime of $\delta t \approx 10$~Myr \citep{LadaLada2003, weidner+2004}, a
molecular cloud will form several embedded clusters whose masses are distributed
according to the \emph{embedded~cluster~mass~function} (ECMF). In galaxies with
ongoing star~formation, many molecular clouds coexist within each $\delta t$
interval. Their combined sIMFs gives rise to a \emph{galaxy-wide~IMF} (gwIMF).
A conceptual illustration is shown in  \cite{jerabkova+2018}, their Fig.~1. See also \citet{Jerabkova+2025}.

If the sIMF were a scale-free probability density distribution function (PDF,
e.g., \citealt{elmegreen1997, elmegreen2006, kj21}), summing sIMFs from multiple
star-forming regions would consistently produce the same composite sIMF,
differing only by Poisson variations and normalizations \citep{yan+23}. In
this case, the gwIMF would have the same shape as the sIMF. This is a common
approach when modeling galaxies (e.g. \citealt{borrow+2022}). There are,
however, arguments indicating that the gwIMF does not always coincide with individual sIMFs
\citep{kroupa+2024}. 

One of the most mature gwIMF models grounded on the physical properties of galaxies is the ``integrated
galactic initial mass function''  (IGIMF) theory, whose historical development and observational evidence is presented in \citet{kroupa+2013, kroupa+2024}. Based on
observations of the sIMF variability \citep{kroupa2002}, \citet{kroupa03}
proposed to calculate the gwIMF as the cumulative sum of all sIMF forming within
a time interval of $\delta t$.

The IGIMF is founded on the following  principles, further motivated in Sec.~\ref{sec:discussion}:
\begin{enumerate}
  \renewcommand{\labelenumi}{(\roman{enumi})}
    \item The gwIMF can be described as the cumulative sum of all sIMFs forming within a galaxy over a short ($\delta t = 10^7$~yr, \citealt{weidner+2004}) time interval, which is the typical lifetime of molecular clouds (Sec.~\ref{sec:intro-foundation}). \label{foundation:1} 
    \item The sIMF, and consequently the gwIMF, are variable. Environmental properties characterizing embedded clusters, such as metallicity and clump gas density, $\rho_{\rm cl}$, affect the resulting shape of the sIMF. We discuss observational evidence for the variability of the sIMF and gwIMF in Sec.~\ref{sec:intro-variation}.
    \item For massive stars to form, a gravitational collapse of dense gas structures must occur \citep[see][ their Sec.~3]{kroupa+2024}. If clumps are not sufficiently massive and dense, they cannot sustain the generation of massive stars. A tight relation arises between the most massive star, $m_{\rm max}$, formed in an embedded clump, and the total stellar mass produced in the embedded clump, $M_{\rm ecl}$ \citep{yan+23}. This $m_{\rm max}-M_{\rm ecl}$~relation is encapsulated mathematically with \emph{optimal~sampling} (Sec.\ref{sec:intro-optimalS}), which is observationally favored over stochastic approaches (Sec.~\ref{sec:intro-stochastic}).
\end{enumerate}
 In brief, optimal sampling (Eq.~\ref{eq:stellar_optimalsampling} and Eq.~\ref{eq:ecmf_optimalsampling} below) is a deterministic method for generating stellar populations from the sIMF. In optimal sampling, the stellar mass budget of an embedded cluster, $M_{\rm ecl}$, determines the maximum stellar mass, $m_{\rm max}$, that can form. A cloud clump would not generate higher mass stars unless lower mass stars can fully populate the sIMF (Gjergo, Zhang \& Kroupa, submitted). 
 This is consistent with low-mass stars first condensing along the molecular filaments. As the clump contracts, the filaments connect at the densest region,  channeling the gas flow onto the more-massive and most-massive stars \citep[][and references therein]{kroupa+2024}. 
 
 Optimal sampling is particularly relevant for small embedded clusters. While, in principle, a $10^3 \,M_{\odot}$ embedded cluster could accommodate the formation of a $100 \, M_{\odot}$ star, this may only occur at the cost of leaving the lower-mass regime underpopulated. This would introduce gaps and Poisson noise that are not observed (expanded further in Sec.~\ref{sec:intro-optimalS}). The mathematical and physical principles of optimal sampling have been developed in Gjergo, Zhang \& Kroupa (submitted).
 
\subsection{Formalisms, nomenclature, and observations} \label{sec:formalisms} 

Note that definitions such as $\rho_{\rm cl}$, $M_{\rm ecl}$, and most quantities in these formalisms are auxiliary states that encapsulate patterns of star formation \citep{kroupa+2013, kroupa+2024}.
The very idea of an IMF is an auxiliary mathematical construct (Hilfskonstrukt in the original German, \citealt{kroupajerabkova2018}): it does not represent any single instantaneous configuration observed in nature. Rather, it is an empirical pattern that serves as a statistical descriptor of stellar populations formed under similar physical conditions. They enable theoretical modeling that lead to a refined interpretation of integrated observations.

The lowercase $m$ is used for stellar masses, while the upper case $M_{\rm ecl}$ is used for embedded cluster masses.  The subscript ``cl'' refers to the progenitor gas strictly involved in the star-formation process, while ``ecl'' refers to the total stellar mass of embedded clusters. 

The terms ``top-heavy'', ``top-light'', ``bottom-heavy'', and ``bottom-light'' have precise definitions \citep[see][]{kroupa+2024}: relative to the solar mass ($1\,\text{M}_{\odot}$), ``top'' (``bottom'') refers to the high-mass (low-mass) end of the sIMF distribution. These terms are defined through comparison with the canonical sIMF \citep{Kroupa2001}, where ``heavy'' (``light'') denotes a number distribution higher (lower) than canonical values. Although originally derived for the solar neighborhood, the canonical sIMF happens to characterize well the typical sIMF of the Milky~Way, hence the designation ``canonical'' (see Eqs.~\ref{eq:canonicalIMF} and \ref{eq:canslopes}).

\subsection{Features and Perks of the IGIMF theory}\label{sec:intro-advantages}

If a molecular cloud clump is not sufficiently massive, its gravitational
potential may be too shallow to induce the dense local overdensities required
for the formation of massive stars \citep[see][]{pak2008, pak2009}. Therefore,
the sIMF is affected by the embedded cluster stellar mass, $M_{\rm ecl}$, or
equivalently by the molecular cloud clump density, $\rho_{\rm cl}$. The sIMF is
also affected by metallicity, because the presence of metals enhances cooling,
which in turn increases the fragmentation of a cloud (see
Sec.~\ref{sec:intro-variation}). Since the sIMF varies, and the gwIMF is
constructed from the ensemble of sIMFs, the gwIMF will also vary
accordingly. 

The IGIMF theory proves to be naturally consistent with observations. For
example, the IGIMF predicts that a low-mass galaxy with a small SFR will have a
top-light gwIMF (\citealt{weidner05}, their Fig.~11,
\citealt{jerabkova+2018,yan20, haslbauer+2024, zonoozi+2025}). This is
consistent with the mass-metallicity relation observed in galaxies
\citep{koeppen+2007, RecchiKroupa2015, yan20, yan+21, haslbauer+2024}, and by
implication, it explains the deficient H$\alpha$- over UV-luminosity observed by
\cite{Lee+2009} for dwarf galaxies \citep{pa+2009}. And indeed, there is direct
evidence for a lack of massive stars in star-forming dwarf galaxies  such as
DDO~154 \citep{watts+2018}. In the starburst environment of 30 Dor in the Large Magellanic Cloud, \cite{Schneider2018} found an excess of massive stars compared to a canonical IMF by counting the stars.

In the context of star-forming dwarf galaxies, the IGIMF theory leads to a significantly higher SFR than what is derived through an invariant sIMF. This results in better agreement with the stellar mass build-up in dwarf galaxies, which may form with nearly constant SFRs in less than a Hubble time \citep{pak2009}. Observationally, the low stellar output of these systems has often been interpreted as evidence for low star-formation efficiency \citep{Shi2014}, which is equivalent to an sIMF variation producing too few massive stars.

On the higher end of the SFR spectrum, \citet{gunawardhana+2011} noted with H$\alpha$ and UV observations that massive star-forming disk galaxies have top-heavy gwIMFs. For extreme cases of dusty gas-rich starburst conditions,  \cite{Zhang18} found that only a top-heavy gwIMF can explain the observed isotopic abundance ratios of $^{13}$CO and C$^{18}$O. 
This can be further extended to the so-called `main-sequence' galaxies in the early universe, which have relatively high star formation rates \citep{Guo24}.
These are accounted for naturally by the IGIMF theory \citep{weidner13, yan+2017, jerabkova+2018}.

Simulations of star-forming tidal-dwarf galaxies produce larger stellar masses when the IGIMF is applied, since low SFRs permit substantially more stellar mass build-up than under an invariant gwIMF. In the latter case, massive stars would form even at low SFRs, potentially disrupting the dwarf galaxy through feedback given its shallow gravitational potential \citep{ploeckinger+2014}.
Along similar lines, hydrodynamical simulations of star-forming dwarf galaxies by \cite{SteyrleithnerHensler2023} show that the gwIMF calculated from the IGIMF leads to oscillations in the SFR with periods comparable to the free-fall time of the cloud. The IGIMF-based SFRs are, on average, higher than those obtained using an invariant gwIMF that permits a constant fraction of massive stars. The model galaxy also experiences distinct chemical evolution patterns in the two scenarios. 

The IGIMF is also consistent with properties of axisymmetric disk galaxies, whose extended UV disk is larger than the H$\alpha$-emitting disk \citep{pak2008}. While all stars emit in the UV, only massive stars contribute to H$\alpha$ emission. In an exponential galactic disk, only the inner regions reach surface gas densities high enough to sustain the formation of massive stars. Low-mass star formation can still occur at larger galactocentric radii in smaller embedded clusters. This has also been seen from maser studies in our Milky Way, where \cite{Sun2018} found
a lack of masers associated with massive stars in the outer Galactic disk compared to that in the inner disk. In these outer regions, however, star formation is H$\alpha$-dark meaning there O-type stars are not produced (other SFR tracers may be missing as well, see also Sec.~\ref{sec:thegwIMF}).

The reported top-heavy gwIMFs in starbursts galaxies at redshifts $2<z<3$ are also consistent with the theoretical expectations of the IGIMF \citep{ZY+2018}. It therefore comes as no surprise that the IGIMF theory, which implies a SFR- and $Z$-dependent gwIMF shape, accounts well for the properties of elliptical galaxies which needed to have formed with top-heavy gwIMFs \citep{yan+21}: the metallicity and photometric properties of the stellar populations in massive elliptical galaxies considered together imply that said stellar populations must have formed very rapidly \citep[within 1 Gyr, ][and references therein]{eappen+2022} with SFRs between $10^3$ and $10^4\,\text{M}_{\odot}/$yr and very early in cosmic history, with negligible star formation at later ages \citep[e.g.,][]{salvador-rusinol+2020} and with major implications for the cosmic microwave background \citep{gjergokroupa25}.
Furthermore, under the IGIMF theory, the gwIMF of an elliptical galaxy at supersolar metallicity will be very bottom-heavy. Strikingly, the masses of super-massive black holes and their scaling with host galaxy mass emerge naturally within the IGIMF theory \citep{kroupa+2020}.

\subsection{This work}\label{sec:intro-ourwork}

In this work, we extend the analysis to supersolar metallicities up to 10~$Z_{\odot}$. We compare the behavior of the sIMF and gwIMF within the IGIMF framework in this regime. We find that the production~rate of low-mass stars increases with metallicity. In other words, the sIMF becomes increasingly bottom-heavy.

The low-mass (bottom) end of the sIMF is highly sensitive to metallicity. We consider the possibility that the current linear dependence of the low-mass sIMF slope, $\alpha_1(Z)$, has a linear dependence on metallicity and can be extrapolated to extreme values. With the existing literature, estimates of the $\alpha_1(Z)$ slope can be inferred for systems of up to $Z=2\,Z_{\odot}$ (i.e., for metal mass fractions of $Z\approx 0.3$ and metallicities $[Z] \approx 0.3$, see Appendix~\ref{sec:metallicity} for definitions). 
At these metallicities, the dependence of fragmentation on metal content may break down. A transition could then occur to a metallicity-insensitive regime in which $\alpha_1(Z)$ no longer increases with metallicity.

We provide an easy-to-use open-source Python package\footnote{\url{https://github.com/egjergo/pyIGIMF}} that quickly calculates stellar populations of galaxies.\footnote{A related open-source python package, {\tt GalIMF}, designed for chemical-evolution studies within the IGIMF theory, has also been made available by \cite{yan+2017}. \url{https://github.com/Azeret/galIMF}.
An extension allowing the calculation of photometric properties of the galaxies is available as the Python code {\tt photGalIMF} \citep{haslbauer+2024}, and the FORTRAN code {\tt SPS-VarIMF} allowing the computation of spectra and photometry of galaxies has been made available by \citep{zonoozi+2025}.}
The new \pyIGIMF\, package is applied here to predict the $m_{\rm max}$–$M_{\rm ecl}$ relation for galaxies of different metallicity. At present, this relation is constrained only near solar metallicity. But since metal-poor and metal-rich gas of the same density exhibit different cooling rates and fragmentation behavior, and because stellar feedback depends on $Z$, the relation is expected to shift with metallicity.

This shift can be computed under the hypothesis that the sIMF is a $\rho_{\rm cl}$- and $Z$-dependent \emph{optimally~sampled} distribution function (ODF). At sub-solar metallicity, the sIMF has been found to be top-heavy (e.g. \citealt{bk2012, kalari+2018, schneider+2018, yasui22}) such that the $m_{\rm max}-M_{\rm ecl}$ relation should shift to smaller $M_{\rm ecl}$ for a given $m_{\rm max}$. The currently known metal-dependence of the sIMF suggests that the inner regions of massive elliptical galaxies that are at supersolar-abundance ought to have very bottom-heavy gwIMFs, as has been argued  from spectroscopic analysis (e.g. \citealt{vDC2010, salvador-rusinol+2021, vDC2021}). Such a significantly bottom-heavy sIMF will lead to a significant shift in the $m_{\rm max}-M_{\rm ecl}$ relation for metal-rich very young star clusters towards large $M_{\rm ecl}$ values that can be observationally tested for. We emphasize that if and only if the sIMF is a systematically variable ODF, then the $m_{\rm max}-M_{\rm ecl}$ relation will shift as described above and quantified below.

In \mysec \ref{sec:methods} we overview the theoretical framework of the IGIMF, which includes both the distribution of the embedded cluster mass function and of the sIMF. We also present an alternative prescription for the low-mass end of the sIMF. In \mysec \ref{sec:results} we present variations of the sIMF over a wide range of physical conditions, and we also quantify the shift of the $m_{\rm max}-M_{\rm ecl}$ relation with metallicity and cloud core density. 
Other approaches formulating the IMF on the galaxy-wide scale have recently been proposed, and we discuss these in comparison to the IGIMF theory in Sec.~\ref{sec:othergwIMFs}. 
A historical overview of the development of the IGIMF theory, along with its observational support, is presented in Sec.~\ref{sec:discussion}.

\section{Methods}\label{sec:methods}

In this section, we first start from the concept of the canonical IMF (Sec.~\ref{sec:canonicalIMF}). We present the mathematical formulation of the IGIMF theory, which is given by the cumulative sum of sIMFs (Sec.~\ref{sec:sIMF}) relative to the embedded cluster mass distribution (Sec.~\ref{sec:ECMF}). To derive the gwIMF, we integrate the sIMFs over the ECMF distribution (as explained in Sec.~\ref{sec:IGIMF}). 
We then present an alternative prescription (Sec.~\ref{sec:logistic}) for the low-mass end of the sIMF slope, $\alpha_1$. Interestingly, the $\alpha_1$ prescription affects the relation between the most massive star in an embedded cluster and the total stellar mass of the cluster itself ($m_{\rm max}$-$M_{\rm ecl}$ relation) for high metallicities.

\subsection{The Canonical IMF}\label{sec:canonicalIMF}
The concept of the IMF as the distribution of initial stellar masses
was introduced by \citet[][S55]{Salpeter1955} (for an overview, see \citealt{kroupa+2024}, their Sec.~1), who proposed a single-slope power~law:
\begin{equation}\label{eq:salpeter}
\xi_{\rm S55} = \frac{\diff N}{\diff m} = k \, m^{-\alpha_{\rm S55}} \, ,
\end{equation} 
where $N$ is the number of stars, $m$ is the stellar mass, $k$ is a normalization constant, and $\alpha_{\rm S55} = 2.35$ is the Salpeter slope (see also \citealt{kroupajerabkova2019} for an historical overview of Salpeter's contribution). 
The sIMF is typically 
 normalized as a probability distribution function, such that the integral of $\xi_{\rm S55}(m)$ over the full mass range is unity.

\begin{figure}
    \centering
\includegraphics[width=\columnwidth]{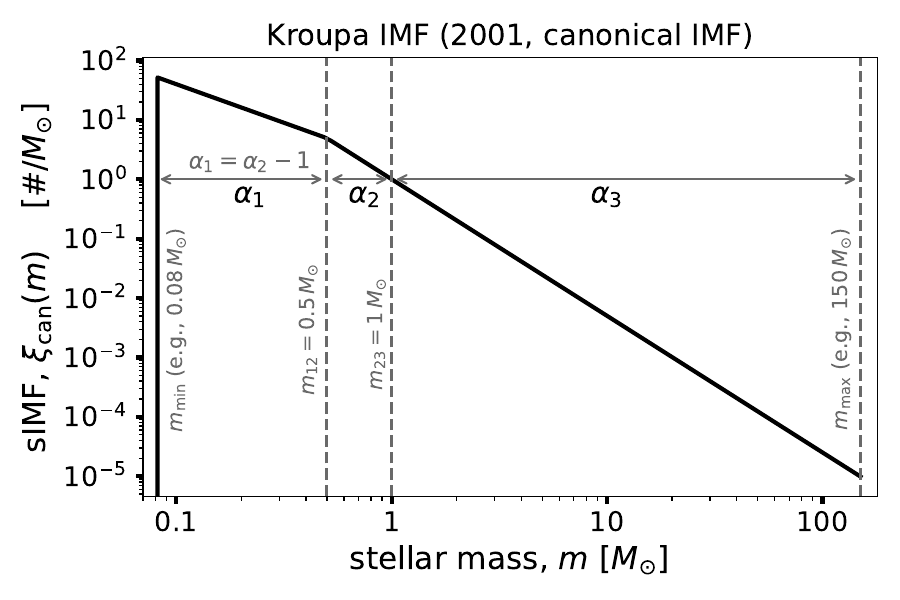}
    \caption{Canonical sIMF \citep{Kroupa2001}, Eq.~\ref{eq:canonicalIMF} and Eq.~\ref{eq:canslopes}, representative of the Milky-Way average and in studies that treat the sIMF as an invariant distribution, often called  ``\emph{universal IMF}''.
    }
    \label{fig:canonicalIMF}
\end{figure}

\citet{Kroupa2001} drew on detailed studies of the solar neighborhood, including nearby star-forming regions and open clusters, and anchored theoretical IMF models in resolved stellar populations rather than integrated-light assumptions, as in some earlier works. The empirical features of the resulting IMF are well described by a broken power law:
\begin{equation}\label{eq:canonicalIMF}
    \xi_{\rm can} = k_i \, m^{-\alpha_{\rm i,can}}  \, ,
\end{equation}
where the slopes, $\alpha_i$, are defined according to the mass interval:
\begin{align}\label{eq:canslopes}
\alpha_{\rm 1,can} &= 1.3 &&\text{for } 0.08 \leq m/\text{M}_{\odot} < 0.5 \nonumber \\
\alpha_{\rm 2,can} &= 2.3 &&\text{for } 0.5 \leq m/\text{M}_{\odot} < 1.0 \nonumber \\
\alpha_{\rm 3,can} &= 2.3 &&\text{for } 1.0 \leq m/\text{M}_{\odot} \, .
\end{align}
The turnover point at 0.5 solar masses reflects the distribution of the stellar masses, as shown in \citet{Kroupa2001}. It is mathematically convenient, implying a fixed relationship of the normalization constants: $k_3 = k_2 = \frac{1}{2} k_1$, without requiring additional continuity conditions.

The above  IMF (Eq.~\ref{eq:canonicalIMF}~along with~\ref{eq:canslopes}) is also known as the ``canonical'' IMF because, while being derived in the solar neighborhood, it also applies to the whole Galaxy, as well as to most Milk-Way-class spiral galaxies. Given that spiral galaxies constitute the most commonly observed morphology, works involving large cosmic volumes may use the canonical IMF as a ``universal'' IMF. However, the earlier notion of an invariant IMF has long been superseded. Andrew Hopkins remarked at the ESO/GALCROSS conference (2024, Brno, Czech Republic) that the question is no longer whether the IMF varies, but rather by how much.

Using the detailed analysis by \citet{Kroupa2001}, \citet{chabrier2003} replaced the segmented power-law slopes $\alpha_1$ and $\alpha_2$ of the low-mass regimes with a smooth lognormal function. This modification yields results that are observationally indistinguishable from \citet{Kroupa2001}, and both formulations are suitable and equivalent representations of the canonical IMF. Some authors, when referring to the ``Kroupa IMF'' actually intend the \citet{kroupa+1993} formulation, where the high-mass slope, $\alpha_3=2.7$, was based on the \citet{scalo86} analysis of massive stars in the nearby Galactic field, out to a few kpc.

\citet{kroupa2005, kroupa+2013} and \citet{Hopkins2018} emphasized the importance of distinguishing between the stellar initial mass function, sIMF, defined for individual star-forming regions (i.e., embedded clusters) and the galaxy-wide IMF, gwIMF, that characterizes an entire galaxy. Throughout this manuscript, we explicitly differentiate between these two forms and avoid using the ambiguous acronym ``IMF''.

Fig.~\ref{fig:canonicalIMF} summarizes the features of the canonical sIMF as it was formulated in \citet{Kroupa2001}. It fits well Milky-Way-like spiral galaxies because of the prevalence of moderate SFRs (a few $\text{M}_{\odot}$/yr) at around solar metallicities in this morphology. 
 The sIMF is defined from 0.08~$\text{M}_{\odot}$ up to a theoretical upper limit of 150~$\text{M}_{\odot}$.  
 The true upper stellar mass in a given embedded cluster may fall below the theoretical maximum mass. 
 Objects less massive than 0.07--0.08~$\text{M}_{\odot}$ are brown dwarfs. They have a distinct origin and are characterized by their own IMF (\citealt{thies+2015} and \citealt{kroupa+2024}, see their Fig.~7).

\subsection{The stellar initial mass function}\label{sec:sIMF}

The sIMF for individual embedded clusters,
$\xi_{\star} (m | M_{\rm ecl}, Z) =\mathrm{d} N/\mathrm{d} m$ is a function of stellar mass, $m$, and will vary in shape according to the embedded cluster mass, $M_{\rm ecl}$, as well as metal mass fraction, $Z$.
$\xi_{\star} (m| M_{\rm ecl}, Z)$ is defined as \citep{Kroupa2001, kroupa+2024}:
\begin{equation}\label{eq:sIMF}
    \xi_{\star}=   k_\star 
    \left\{ \begin{array}{ll}
    \,\,\,\,\,\,\,\, \, m^{-\alpha_1(Z)}, \hspace{1.1cm} m_{\rm min}\leq {m} <m_{\rm 12} \,, \\
    \,\, k_{\mathrm{2}} \, m^{-\alpha_2(Z)}, \hspace{1.3cm} m_{\rm 12}\leq {m} <m_{\rm 23} \,, \\
    \,\, k_{\mathrm{3}} \, m^{-\alpha_3(Z, M_{\rm ecl})}, \hspace{0.65cm} m_{\rm 23}\leq {m} < m_{\mathrm{max}} \,, \\
    \end{array} \right. \, ,
\end{equation}
where the lower mass limit is set to $m_{\rm min} = 0.08 \, \text{M}_{\odot}$, $k_{\rm \star}$ is the normalization constant, and $k_{\rm 2}, k_{\rm 3}$ ensure continuity of the power~law segments.

A functional dependence of the sIMF power-law indices on the iron abundance, $[\rm Fe/H]$ (Eq.~\ref{eq:MvHbymass}), was first introduced by \citet{kroupa2002} and then applied by \citet{marks+12}. The modern formulation by \citet{yan+21}, which uses the metal mass fraction, $Z$ (Eq.~\ref{eq:metalmassfrac}), is encapsulated as follows:
\begin{equation}\label{eq:alpha1}
\begin{split}
    &\alpha_1=1.3+\Delta\alpha \cdot (Z-Z_\odot),\\
    &\alpha_2= \alpha_1 + 1,
\end{split}
\end{equation}
where $Z_{\odot}=0.0142$ refers to the metal mass fraction, from \citet{asplund09}. \citet{yan+21} found the fit for $\Delta \alpha$ to be $\Delta \alpha = 63$.

The high-mass sIMF slope, $\alpha_3$ depends weakly on the metallicity, and more strongly on the gas clump density, $\rho_{\rm cl}$. Since $\rho_{\rm cl}$ characterizes the conditions during the protostellar phase, it effectively encapsulates both the initial gas mass and the resulting stellar mass, $M_{\rm ecl}$, of the embedded cluster. \citet{marks+12} and then \citet{yan+21} obtain the following best fits:
\begin{equation}\label{eq:alpha3}
\alpha_3=
        \begin{cases} 
            2.3, & x<-0.87 \, , \\
            -0.41x+1.94, & x>-0.87 \, ,
        \end{cases}
\end{equation}
where $x$ is described by:
\begin{equation}\label{eq:x_rhocl}
    x=-0.14 \, [\mathrm{Z}]+0.99 \, \log_{10}(\rho_{\mathrm{cl}}) - 6 \, ,
\end{equation}

\noindent
where the $-6$ comes from normalizing $\rho_{\rm cl}$ inside of the log by $10^6$. These relations make the sensible assumption (\citealt{kroupa+2024} and references therein) that the star-formation efficiency of a clump is 33~per cent ($\frac{1}{3}$ of the gas clump mass is condensed into stars), leading to the embedded clusters follow a radius--mass relation:
\begin{equation}\label{eq:rho_cl}
\log_{10}\rho_{\mathrm{cl}}=0.61\log_{10}M_{\rm ecl}+2.85 \, .
\end{equation}
The above equation assumes astronomical units, i.e., $\text{M}_{\odot}$ for $M_{\rm ecl}$ and $\rm M_{ \odot} pc^{-3}$ for $\rho_{\rm cl}$ (see also Sec.~\ref{sec:intro-variation} and Sec.~\ref{sec:formalisms}).

\begin{figure}
    \centering
    \includegraphics[width=\columnwidth]{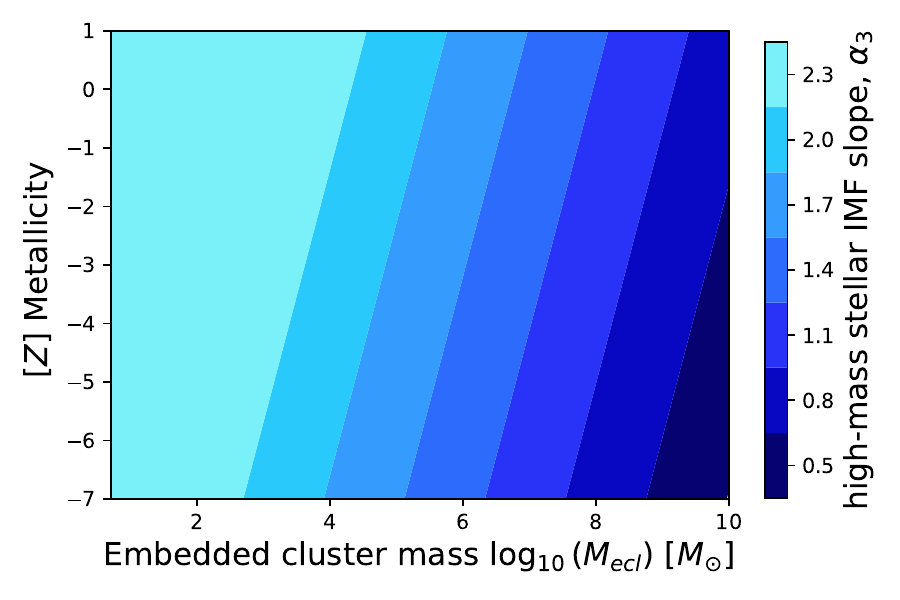}
    \caption{Values spanned by the sIMF slope for massive ($> 1 \text{M}_{\odot}$) stars, $\alpha_3$ (Eq.~\ref{eq:alpha3}), within the possible gas densities and metallicities covered  by the embedded cluster. 
     }
    \label{fig:alpha3}
\end{figure}

Fig.~\ref{fig:alpha3} shows the range of $\alpha_3$ values as a function of its physical parameters, namely metallicity and $M_{\rm ecl}$. 
Note that there is no clear evidence for sIMFs with $\alpha_3>2.3$ \citep{kroupa+2024}.
The steeper Galactic-field or Solar neighborhood IMFs  for stars ($m > 1 \, \text{M}_{\odot}$, with $\alpha_3\approx 2.7$, \citealt{scalo86, kroupa+1993, kroupa+2024} and references therein) result from composite IMFs, i.e., field stars originate from many embedded clusters. The region with $M_{\rm ecl} > 10^8\, M_{\rm ecl}$ and $[Z]<-3$ is relevant for the formation of supermassive black holes \citep{kroupa+2020}.

\begin{figure*}[htbp]
    \centering
    \includegraphics[width=\columnwidth]{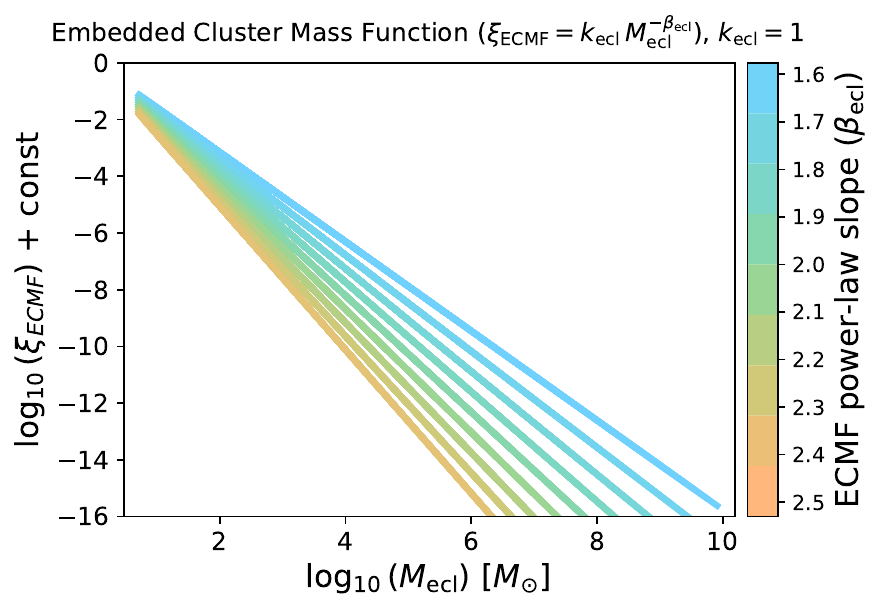}
    \includegraphics[width=\columnwidth]{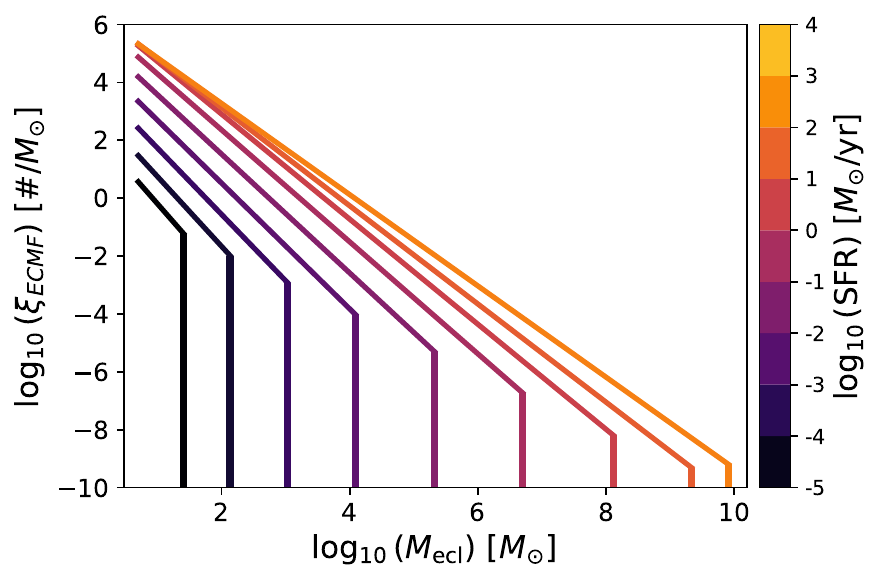} 
    \caption{(\emph{Left:}) The embedded cluster mass function (ECMF,  $\xi_{\rm ECMF}$, Eq.~\ref{eq:xi_ecl}) shown as an unbounded power law, normalized to  $k_{\rm ecl} = 1$. The slope of the power law,  $\beta_{\rm ecl}$ (Eq.~\ref{eq:beta-SFR}), depends on the SFR and is indicated by the color-bar. Steeper ECMFs correspond to higher  $\beta_{\rm ecl}$, which occur at lower SFRs. This visualization, normalized to $k_{\rm ecl}=1$, facilitates tracking the slope variation. (\emph{Right:}) The ECMFs, normalized and truncated to their respective $k_{\rm ecl}$ and $M_{\rm ecl,max}$, as determined by optimal sampling (as in Fig.~\ref{fig:ECMFparameters}). The SFR is indicated by the color-bar, with 9 discrete values matching those used for  $\beta_{\rm ecl}$  in the left panel. At lower SFRs, ECMFs are steeper and limited to smaller  $M_{\rm ecl,max}$, whereas higher SFRs allow for more massive clusters to form. At the typical SFR of main-sequence galaxies, a few $\text{M}_{\odot}/{\rm yr}$, the most massive embedded cluster that may form is approximately  $10^6 \, \text{M}_{\odot}$.}
    \label{fig:ECMFvariation}
\end{figure*}

To normalize the sIMF so that its integral over the full stellar mass range correctly identifies the total number of stars produced, we invoke optimal sapling (\citealt{kroupa+2013, schulz15}, see also Sec.~\ref{sec:intro-optimalS}):
\begin{equation}\label{eq:stellar_optimalsampling}
    \left\{
    \begin{aligned}
        M_{\mathrm{ecl}} &= \int_{0.08~\mathrm{M}_{\odot}}^{m_{\mathrm{max}}}m~\xi_{\mathrm{\star}}(m)\,\mathrm{d}m \,, \\
        1 &= \int_{m_{\mathrm{max}}}^{m_{\rm max *}} \xi_{\star}(m)  \,\mathrm{d}m \, .
    \end{aligned}
    \right.
\end{equation}
where $M_{\rm ecl}$ is the embedded cluster stellar mass and $\xi_{\star}$ is the sIMF for stars of mass $m$.  $m_{\rm max*} \approx 150\,M_\odot$ is the physically-possible upper mass limit to born stars as determined empirically (see \citealt{weidner+2004, Oey2007, zinneckeryorke2007}, \citealt{kroupa+2013}, their Sec.~3.1 and Sec.~3.2, also \citealt{kroupa+2024}, their Sec.~3.3, and references therein).  
Given that stars are discrete entities, for the smallest embedded clusters it is recommended to discretize Eq.~\ref{eq:stellar_optimalsampling} as in \citet{schulz15}. However, as explained in 
Gjergo, Zhang \& Kroupa, (submitted, their appendix), the error may be as small as 1\% for embedded clusters containing over 10 stars.

\subsection{The embedded cluster mass function}\label{sec:ECMF}

\begin{figure}
    \centering
\includegraphics[width=\columnwidth]{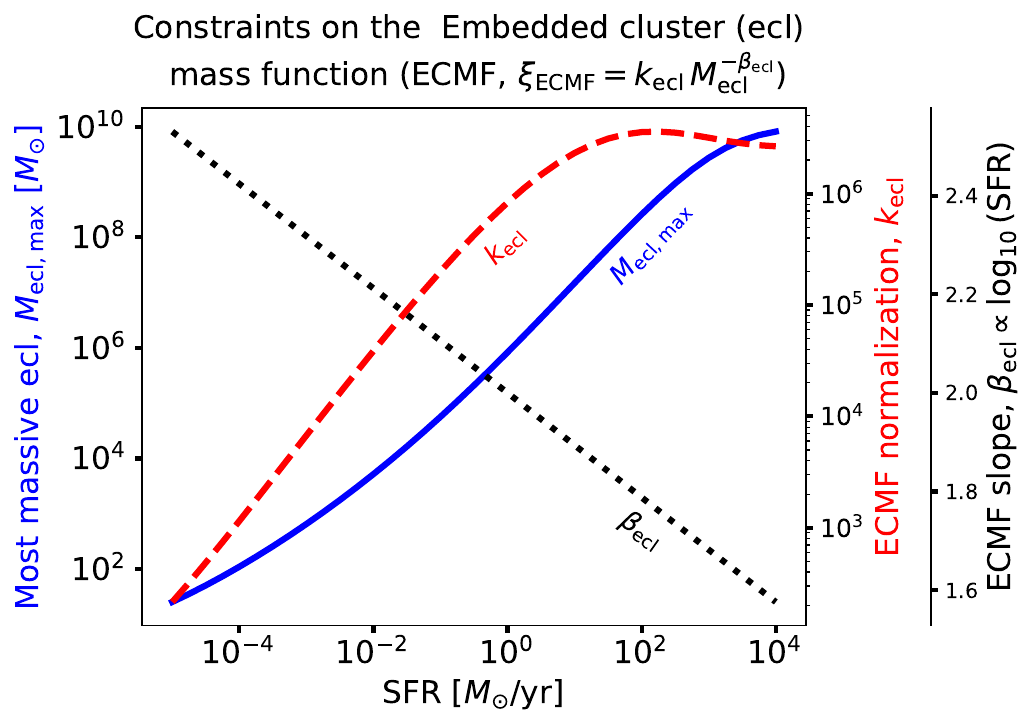}
    \caption{Constraints on the embedded cluster mass function (ECMF) imposed by optimal sampling, where the ECMF describes the mass distribution of embedded clusters formed in a galaxy within a time interval $\delta t$. At any time  $t_i$,  within a time interval of  $\delta t = 10^7$~yr, the total stellar mass produced in a galaxy is given by the top equation in  Eq.~\ref{eq:ecmf_optimalsampling}. The slope of the embedded cluster mass function,  $\beta_{\rm ecl}$ (\emph{black dotted curve}), depends on the SFR (Eq.~\ref{eq:beta-SFR}). The other two parameters, determined by optimal sampling, are: the most massive embedded cluster that may form at  $t_i$,  $M_{\rm ecl,max}$ (\emph{blue solid curve}), and the normalization of the embedded cluster mass function,  $k_{\rm ecl}$ (\emph{red dashed curve}).}
    \label{fig:ECMFparameters}
\end{figure}

The embedded cluster mass function (ECMF, $\xi_{\rm ECMF}$) can be represented by a power law (see \citealt{kroupa+2013} and references therein):
\begin{equation}\label{eq:xi_ecl}
\xi_{\mathrm{ECMF}}=\frac{\mathrm{d} N_{\rm ecl}}{\mathrm{d} M_{\rm ecl}}=
k_{\mathrm{ecl}} M_{\rm ecl}^{-\beta_{\rm ecl}}, 
\end{equation}
which is valid in the embedded cluster mass range $5 \text{M}_{\odot} \leqslant M_{\rm ecl} <M_{\mathrm{ecl,max}}$,
\citep{weidner+2004,  km2011, joncour+18}
where the power law index $\beta$ will depend on the SFR.
The empirical dependence of the index of the power law and the average SFR of a galaxy over the time interval $\delta t =10\,$Myr
($\bar{\psi}_{\delta t}$) is given by \citep{yan+21}:
\begin{equation}\label{eq:beta-SFR}
\beta_{\rm ecl}=-0.106\log_{10} \bar{\psi}_{\delta t} +2.
\end{equation}

The left panel of \myfig \ref{fig:ECMFvariation} shows the range spanned by $\beta_{\rm ecl}$ for an average SFR in the range between $10^{-5.5}$ and 10$^4 \text{M}_{\odot}$/yr. The right panel shoes the normalized ECMF.  
While the mass function of very young embedded clusters follows a power law at a given location in a galaxy, the galaxy-wide ECMF takes the form of a Schechter-like function in spiral galaxies. This transition arises naturally from the spatial distribution of mass, and hence the local SFR surface density, across the disk \citep{LieberzKroupa2017}. We defer detailed analysis of the spatially-distributed ECMF to future works.

Embedded clusters (ECs) span a wide range of masses. In the extreme star-burst cases \citep{BK2018}, an EC may produce tens of millions of stars. Low-mass embedded clusters will produce $\approx 50-100\,\text{M}_{\odot}$ of stars, $M_{\rm ecl}$, but will be deprived of massive stars \citep{hsu+2012}. As for the low-mass end, \citet{km2011} and \citet{joncour+18} observed embedded clusters as small as a~few~low-mass binaries, leading to a low-mass estimate of $M_{\rm ecl} \approx 5\,\text{M}_{\odot}$.

To return the correct number of embedded clusters once integrated, the embedded cluster mass from \myeq \ref{eq:xi_ecl} needs to be normalized. This is achieved applying the system of equations from optimal sampling (for a detailed discussion of the method, see Sec.~\ref{sec:intro-optimalS}):
\begin{equation}\label{eq:ecmf_optimalsampling}
    \left\{
    \begin{aligned}
        M_{\rm tot} =\bar{\psi}(t) \, \delta t &= \int_{M_{\rm ecl, min}}^{M_{\rm ecl,max}} M~\xi_{\mathrm{ecl}}(M)\,\mathrm{d} M \,, \\
        1 &= \int_{M_{\mathrm{ecl,max}}}^{M_{\rm ecl, max *}} \xi_{\rm ecl}(M)  \,\mathrm{d}M \, .
    \end{aligned}
    \right.
\end{equation}
where $\bar{\psi}(t)$ is the average star~formation~rate sampled over a time interval $\delta t = 10^7 \, {\rm yr}$ around time $t$. and $\xi_{\rm ecl}$ is the embedded cluster mass function for embedded clusters of mass $M$ (for compactness, we use $M = M_{\rm ecl}$). $M_{\rm ecl,max *}$ is the theoretical upper mass limit for any embedded cluster and it is taken to be $10^{10} \, \text{M}_{\odot}$. This corresponds to the birth mass of an ultra-compact dwarf galaxy \citep{dabringhausen+09, dabringhausen+12, jerabkova+2018, mahani+21, Zonoozi+16, Haghi+17}.

In practice, we find $M_{\rm ecl, max}$ and $k_{\rm ecl}$ by taking the ratio between the two equations in the system above (Eq.~\ref{eq:ecmf_optimalsampling}), which cancels out the presence of the embedded cluster mass function normalization constant, $k_{\rm ecl}$, and we find the root of:
\begin{equation}\label{eq:norm1ECMF}
    \bar{\psi}(t) \, \delta t \, \int_{M_{\mathrm{ecl, max}}}^{M_{\rm ecl, max *}} M^{-\beta}  \,\mathrm{d}M  - \int^{M_{\mathrm{ecl, max}}}_{M_{\rm ecl,min}} M^{1 -\beta}\,\mathrm{d}M = 0 \, ,
\end{equation}
which returns the best fit for $M_{\rm max}$, the real most massive embedded cluster formed in a galaxy at time $t$. $k_{\rm ecl}$ is then obtained substituting $M_{\rm max}$ back into Eq.~\ref{eq:ecmf_optimalsampling}. This procedure is equivalent to optimal sampling used for the sIMF (Eq.~\ref{eq:stellar_optimalsampling}).

Fig.~\ref{fig:ECMFvariation} displays the ECMF. On the left-hand side, the normalization constant is set to $k_{\rm ecl}=1$. On the right-hand side, the ECMF is normalized according to optimal sampling. The vertical lines on the right-hand side represent the effective upper mass limit for each cluster at a given SFR. Notice that the slopes between the two figures are the same, only the $k_{\rm ecl}$ varies. 
Fig.~\ref{fig:ECMFparameters} depicts the three parameters involved in the ECMF normalization, i.e.,  the SFR-dependent power-law slope, $\beta_{\rm ecl}$, and the two quantities derived from solving the system of equations under optimal sampling (Eq.~\ref{eq:ecmf_optimalsampling}), the most massive embedded cluster stellar mass, $M_{\rm ecl, max}$, and the normalization constant, $k_{\rm ecl}$. The normalization $k_{\rm ecl}$ increases sharply with SFR, peaking around $10^2 M_{\rm \odot} \text{yr}^{-1}$, and then decreases slightly as $M_{\rm ecl, max}$ approaches its theoretical upper limit and the smaller $\beta_{\rm ecl}$ flattens the ECMF. This behavior implies that at the highest SFRs, the number of low-mass embedded clusters grows only slightly, while the rise in SFR is predominantly accommodated by a steep increase in the formation of high-mass clusters. This behavior is also evident in the right-hand panel of  Fig.~\ref{fig:ECMFvariation}).

\subsection{The integrated galaxy initial mass function}\label{sec:IGIMF}

The gwIMF, just like the sIMF, returns the number distribution of stars over their stellar masses: $\xi_{\rm IGIMF}(m\, |\, \bar{\psi}(t), Z(t)) = \mathrm{d} N/\mathrm{d} m$.
In the IGIMF formulation, the number distribution of stars is integrated across all the galaxy, and across all embedded clusters generated during the time interval  $t$ to $t+\delta t$ and is given by:
\begin{align}\label{eq:IGIMF}
\xi_{\rm IGIMF}(m\, |\,  \bar{\psi}, Z)=&\int_{5 \text{M}_{\odot}}^{M_{\rm ecl,max}}
\xi_{\mathrm{ecl}}(M_{\rm ecl}|\bar{\psi}_{\delta t}) \times\, \nonumber\\
&\xi_{\mathrm{\star}}(m|M_{\rm ecl},Z) \, \mathrm{d}M_{\rm ecl} \, .
\end{align}
The parameters of the IGIMF theory are summarized in Table~\ref{tab:parameters}. The integral in Eq.~\ref{eq:IGIMF} yields a function of stellar mass, $m$, which depends on both the metallicity and the average SFR of a galaxy over a time interval $\delta t$. At any given time $t$, a galaxy is characterized by an average $SFR(t)$, $\bar{\psi}(t)$, and metal~mass~fraction, $Z$, and each interval $t_i$ has an associated $\xi_{\rm IGIMF}$.

\begin{table}[t]
    \centering
    \begin{tabular}{c|c|c}
 Parameter & Value & Meaning \\ \hline
 &&\\
$Z_{\odot}$ & 0.0142 & solar metal mass fraction \\
$\delta t$ & $10^7 \, \mathrm{yr}$ & lifetime of molecular clouds\\ && serving as \\ && star-formation timestep \\ 
$\alpha_{\rm 1,canon}$ & 2.3 & canonical low-mass slope\\ 
$\alpha_{\rm 3,canon}$ & 1.3 & canonical high-mass slope\\ &&\\
$M_{\rm ecl,min}$ & $5 \, \text{M}_{\odot} $ & lower limit for the\\
&& embedded cluster stellar mass \\
$M_{\rm ecl,max *}$ & $10^{10} \, \text{M}_{\odot}$ & theoretical upper limit for the\\
&& embedded cluster stellar mass \\
&& \\
$m_{\rm min}$ & $0.08 \, \text{M}_{\odot}$ &  lower limit for the stellar\\
&& mass in an embedded cluster \\
$m_{\rm max *}$ & $150  \, \text{M}_{\odot}$ & theoretical upper limit for the \\
& & stellar mass in an embedded cluster \\ && \\
$\Delta \alpha$ & 63  & linear $\alpha_1(Z)$ slope \\
$x_0$ & $Z_{\odot}$ & logistic $\alpha_1(Z)$ midpoint \\
$\alpha_{1, \rm max}$ & $2 \, \alpha_{1, \text{canon}}$ &  logistic $\alpha_1(Z)$  maximum \\
$k_b$ & $2/Z_{\odot}$ & logistic $\alpha_1(Z)$ growth rate \\ &&\\
\end{tabular}
    \caption{Parameters assumed in the IGIMF theory and in \pyIGIMF. 
    $Z_{\odot}$ is the solar metal mass fraction from \citet{asplund09} (their proto-solar metallicity from Table 4).
    $\delta t$ is the timestep of star formation for a galaxy, i.e., for the duration of $\delta t$ we can take the SFR to be constant. The value is taken from \citet{schulz15} and corresponds to the effective, i.e. star-contributing, life-time of molecular clouds.
    $M_{\rm ecl,min}$ is the lower mass limit of an embedded cluster total stellar mass. $M_{\rm ecl,max *}$ is the theoretical upper mass limit for an embedded cluster total stellar mass. Note that generally, for a given SFR, the true upper limit, $M_{\rm ecl, max}$, will be lower than the theoretical $M_{\rm ecl, max *}$. $m_{\rm min}$ is the lower mass limit for stars in an embedded cluster. $m_{\rm max *}$ is the theoretical upper mass limit for stars in an embedded cluster. Also in the case of stars, the actual upper mass limit, $m_{\rm max}$, will be smaller than the theoretical upper limit for most low-to-moderate embedded cluster masses. 
    $\Delta \alpha$ is the metallicity slope for the linear prescription of $\alpha_1(Z)$ (Eq.~\ref{eq:alpha1}). $x_0$, $\alpha_{1\rm, max}$, and $k_b$ are the parameters of the logistic prescription of $\alpha_1(Z)$ (Eq.~\ref{eq:alpha1_alternative} and Eqs.~\ref{eq:logisticparam}). 
    }
    \label{tab:parameters}
\end{table}

\subsection{Alternative prescription of the low-mass (bottom) slope} \label{sec:logistic}

\begin{figure*}[ht]
    \centering
\includegraphics[width=0.8\textwidth]{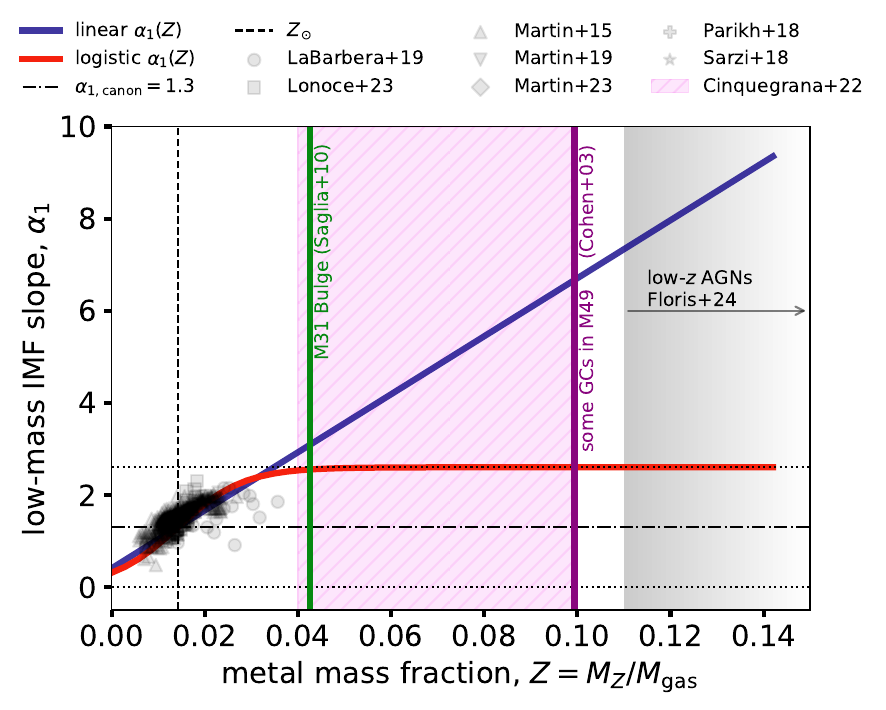}
    \caption{Relation between the low-mass sIMF slope, $\alpha_1$, and metal mass fraction, $Z$, in the linear case (Eq.~\ref{eq:alpha1}, \emph{blue line}) and in the logistic case (Eq.~\ref{eq:alpha1_alternative}, \emph{red line}). The solar metal mass fraction, $Z_{\odot}=0.0142$ \citep{asplund09}, is shown with the vertical \emph{dashed black line}. The canonical IMF slope, $\alpha_{1,\rm canon}$, with the \emph{dash-dot black line}. The \emph{dotted black lines} identify the bounds of the logistic prescription for $\alpha_1(Z)$. The \emph{shaded gray area} is an example of some of the highest metallicities observed in the broad-line regions of AGNs in the local Universe \citep{Floris+2024}.
    However, shear motion around a SMBH can suppress the formation of low-mass stars, despite their high metallicities \citep{kroupa+2024}.
    While these metallicities will be diluted by the time stars form, they offer an empirical indication that such metallicities may be accessible. The shaded pink region refers to theoretical predictions of AGB yields from \citet{Cinquegrana+2022}. The scatter points are inferred from a compilation of observations reported in \citet{Yan+2024}, their Fig.~4, for Galactic Observations. The specific papers that reported the original observations are \citet{LaBarbera+2019, Lonoce+2023, Martin+2015, Martin+2019, Martin+2023, Parikh+2018, Sarzi+2018}. The \emph{green vertical line} is the metal mass fraction of the Andromeda Galaxy Bulge \citep[M31, ][]{Saglia+2010}, while the \emph{purple vertical line} refers to globular clusters in the giant elliptical galaxy M49 \citep[such as NGC~4062 and NGC~5097, ][]{Cohen+2003}. }
    \label{fig:alpha1alternative}
\end{figure*}

Eq.~\ref{eq:alpha1} describes the linear increase of $\alpha_1$ with metallicity, reflecting the role of metals in enhancing cooling and accelerating gas fragmentation. But this enhancement may saturate above a critical (yet unknown) metallicity. 
In this scenario, there is a transition from a regime sensitive to the metallicity variation, to a metallicity-insensitive regime. 
A logistic function is the appropriate formulation to capture the transition between these two regimes.  
We therefore propose that $\alpha_1(Z)$ can be appropriately described by:
\begin{equation}\label{eq:alpha1_alternative}
    \alpha_1(Z)=\frac{\alpha_{1,\rm max}}{1+e^{-k_b(Z-x_0)}},
\end{equation}
$x_0$ is the x-axis midpoint of the logistic function. $\alpha_{1,\rm max}$ is the curve's maximum value, and $k$ is the growth rate. 

We choose the following parameters, which represent the extreme example where $\alpha_1$ does not increase beyond the values tested to date. In this case:
\begin{align} \label{eq:logisticparam}
    x_0 & = Z_{\odot} \, , \nonumber \\
    \alpha_{1, \rm max} & = 2 \, \alpha_{1, \text{canon}} \, , \nonumber \\
    k_b & =2/Z_{\odot}
\end{align}
where $\alpha_{1, \text{canon}}$ is defined in Eq.~\ref{eq:canslopes}.

\section{Results}\label{sec:results}
The results are presented in 3 steps. First we examine the variability of the sIMF as a function of metallicity and embedded cluster mass, using both the linear and logistic $\alpha_1$ prescriptions. It also considers the potential dependence of the $m_{\rm max}$–$M_{\rm ecl}$ relation and the sIMF slopes on metallicity. Next, we addresses the variability of the gwIMF and its dependence on metallicity and SFR. Lastly, we compare these results with other models of a variable gwIMF.

\subsection{The sIMF}
\label{sec:thesIMF}

\begin{figure*}[ht]
    \centering        
    \includegraphics[width=\textwidth]{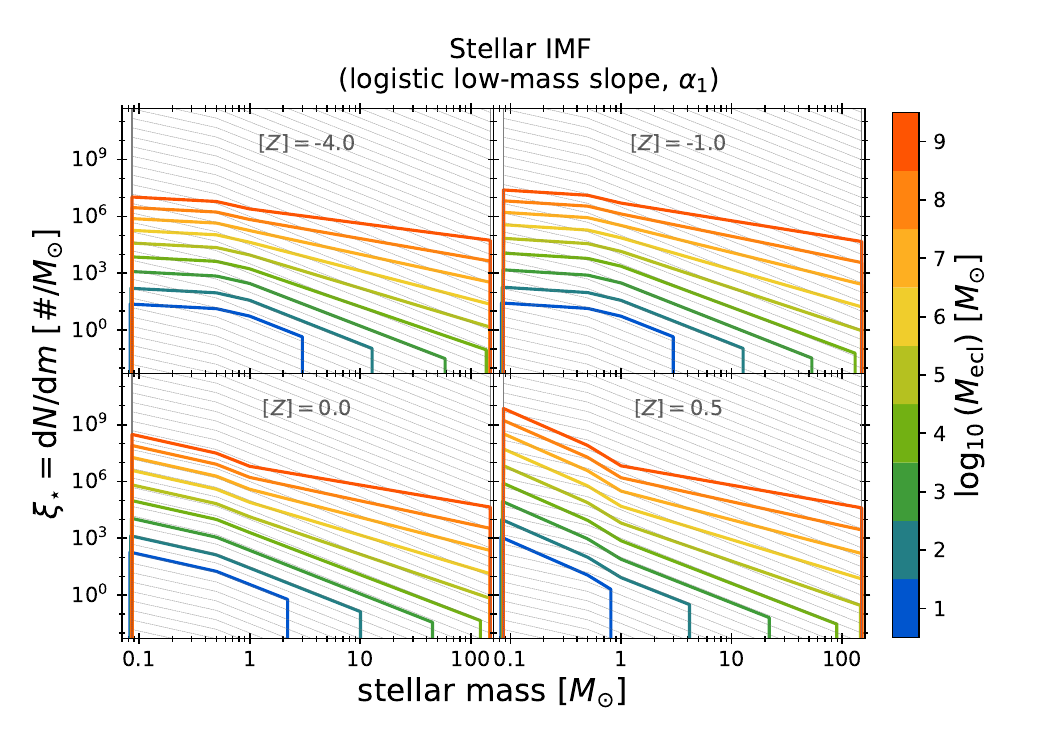}
    \caption{sIMF variations, $\xi_{\star}$ (Eq.~\ref{eq:sIMF}), computed using a logistic $\alpha_1$ dependence on metallicity (Eq.~\ref{eq:alpha1_alternative}) and normalized so that the area under each curve is the number of stars in the embedded cluster. Each panel refers to a different metallicity, $[Z]$, while the color-bar identifies the embedded cluster stellar mass, $M_{\rm ecl}$. The gray lines are parallel to the canonical sIMF (Eq.~\ref{eq:canonicalIMF}). The lower stellar mass limit for all sIMFs is always 0.08~$\text{M}_{\odot}$. For each sIMF, the upper stellar mass limit is constrained by the $m_{\rm max}-M_{\rm ecl}$~relation, which results from optimal sampling (Eq.~\ref{eq:stellar_optimalsampling}).}
    \label{fig:sIMFsbySFR_logistic}
\end{figure*}

Fig.~\ref{fig:alpha1alternative} provides a comparison of $\alpha_1(Z)$ for the linear (Eq.~\ref{eq:alpha1}) and logistic (Eq.~\ref{eq:alpha1_alternative}) formulations  and with observational constraints. A systematic survey investigating the $\alpha_1$ slope with consistent assumptions and methodology does not exist. From existing literature adopting disparate assumptions, \citet{Yan+2024} (e.g., their Fig.~4) compiled mass ratio estimates for stars in the mass range $0.2 < m/\text{M}_{\odot} < 1$:
\begin{equation}
    \xi_{\rm MR} = \frac{\int_{0.2 \text{M}_{\odot}}^{0.5 \text{M}_{\odot}} m \,  \cdot \, m^{-\alpha_1} \diff m}{\int_{0.2 \text{M}_{\odot}}^{0.5 \text{M}_{\odot}} m \,  \cdot \, m^{-\alpha_1} \diff m + \int_{0.5 \text{M}_{\odot}}^{1.0 \text{M}_{\odot}} m \,  \cdot \, m^{-\alpha_2} \diff m} \, .
\end{equation}

We applied a minimization algorithm, specifically Brent's method \citep[][their Sec.~2.8]{Atkinson1989}, to solve for $\alpha_1$, under the assumption that $\alpha_2 = \alpha_1 + 1$ holds at all times. A detailed compilation of the results is provided by \citet{Yan+2024}. IMF slope estimates are available up until a metal mass fraction (Eq.~\ref{eq:metalmassfrac}) of $Z\approx 0.03$, which is approximately 2 times the solar value $Z_{\odot}$ (see Tab.~\ref{tab:parameters}). However, stars and environments with higher supersolar metallicities are known to exist. For example, the Andromeda bulge may contain stars with $[Z]\approx 0.5$ \citep{Saglia+2010}. \citet{Cinquegrana+2022} is among the first theoretical works tackling extreme supersolar metallicities. They computed supersolar AGB yields for metal mass fractions $0.04 < Z < 0.1$. Their paper mentions known stellar populations with supersolar metallicities, but emphasize the scarcity of direct observations due to their rarity in the Galaxy, as well as practical limitations. Therefore there are yet no observational counterparts for AGBs in this metallicity range. However, \citet{Cohen+2003} find that the giant elliptical M49 hosts globular clusters whose metallicities are as high as $Z\approx 7 Z_{\odot}$.

\begin{figure*}[ht]
    \centering
    \includegraphics[width=\textwidth]{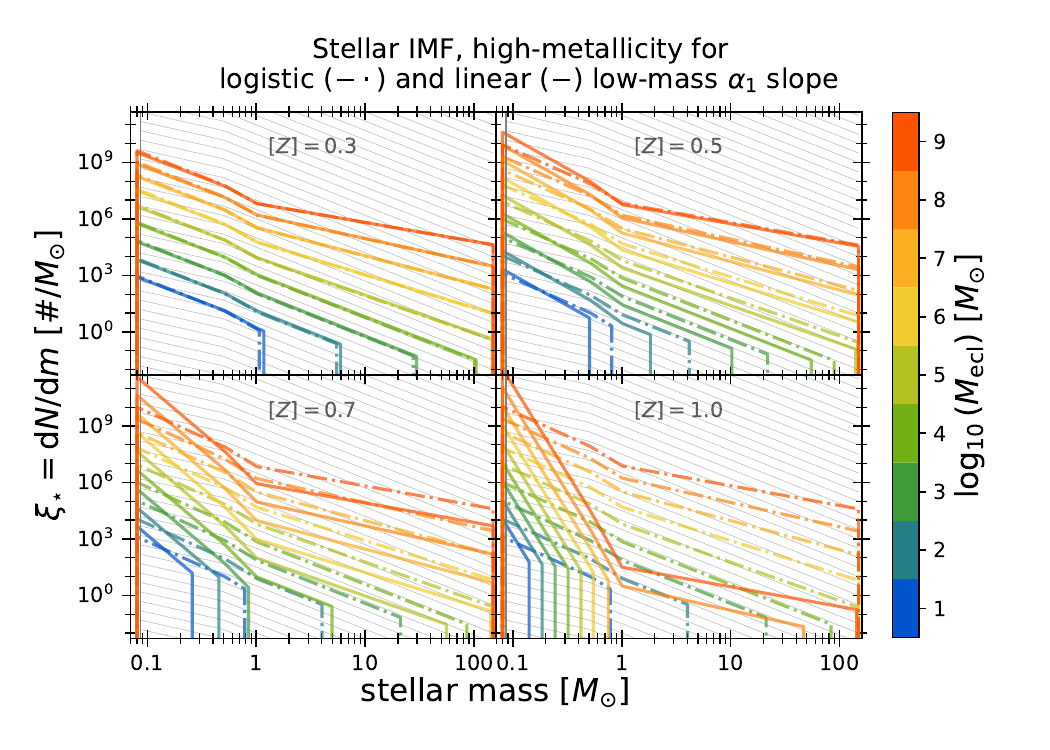}
    \caption{Similar to Fig.~\ref{fig:sIMFsbySFR_logistic},  but for supersolar metallicities, as labeled in each panel. Each color corresponds to a different embedded cluster mass, as indicated by the \emph{color-bar}. The \emph{solid lines} assume a linear dependence of $\alpha_1$ on metallicity (Eq.~\ref{eq:alpha1}), while the \emph{dot-dashed lines} a logistic form (Eq.~\ref{eq:alpha1_alternative}). At $[Z]=1.0$, the logistic prescription allows clusters with $M_{\rm ecl} \gtrsim 10^3 \text{M}_{\odot}$ to produce massive stars ($m> 10 \text{M}_{\odot}$). In contrast, for the same $M_{\rm ecl}$ the linear prescription produces only low-mass stars of around $0.3 \text{M}_{\odot}$. Massive stars are produced only for $M_{\rm ecl} \gtrsim 10^7 \text{M}_{\odot}$ in the linear case. }
    \label{fig:supersolar_sIMF}
\end{figure*}

Recently, \citet{Floris+2024} measured the metallicity in the broad-line regions of local active galactic nuclei (AGNs), adopting the definition in Eq.~\ref{eq:MvHbynumb}, where $[Z/H]$ denotes the number density ratio of all metals to hydrogen. They report among the highest metallicities in the literature, reaching values at least several dozen times solar and up to truly extreme levels of $[Z/H] \approx 3$. It is to be expected that, prior to condensing into molecular clouds, such gas will be diluted by the surrounding metal-poor medium. While these observations do not directly confirm the existence of stellar populations with metallicities as high as $Z = 10 Z_{\odot} = 0.142$, they suggest such values may be possible.

\myfig \ref{fig:sIMFsbySFR_logistic} shows how the sIMF varies as a function of the embedded cluster mass $M_{\rm ecl}$ for different metallicities. $[Z]=0$ is on the bottom left panel. This is the metallicity where the low-mass ($m < 1 \text{M}_{\odot}$) $\alpha_1$ and $\alpha_2$ slopes are the closest to canonical values (Eq.~\ref{eq:canslopes}). Lower metallicities are bottom-light while higher metallicities are bottom-heavy. The high-mass slope, $\alpha_3$, remains canonical as long as the most massive star in the embedded cluster, $m_{\rm max}$, does not approach the theoretical upper limit, $m_{\rm max *}$. Only as $m_{\rm max} \rightarrow m_{\rm max *}$ does the high-mass slope become top-heavy. We stress again that top-light sIMFs are not possible (see the discussion for Fig.~\ref{fig:alpha3} in Sec.~\ref{sec:sIMF}). In Fig.~\ref{fig:sIMFsbySFR_logistic}, we show embedded clusters ranging from $10$ to $10^9 M_{\rm ecl}$. This wide range of values is supported by observations (see Sec.~\ref{sec:intro-foundation} and Sec.~\ref{sec:ECMF}). In the Galactic center, however, the most massive observed embedded clusters reach masses of $\approx 10^4 \text{M}_{\odot}$ as in the case of Westerlund 2 \citep{Zeidler+2021}, corresponding to the bright green curves in Fig.~\ref{fig:sIMFsbySFR_logistic}.

\begin{figure*}[ht]
\centering    
\includegraphics[width=\columnwidth]{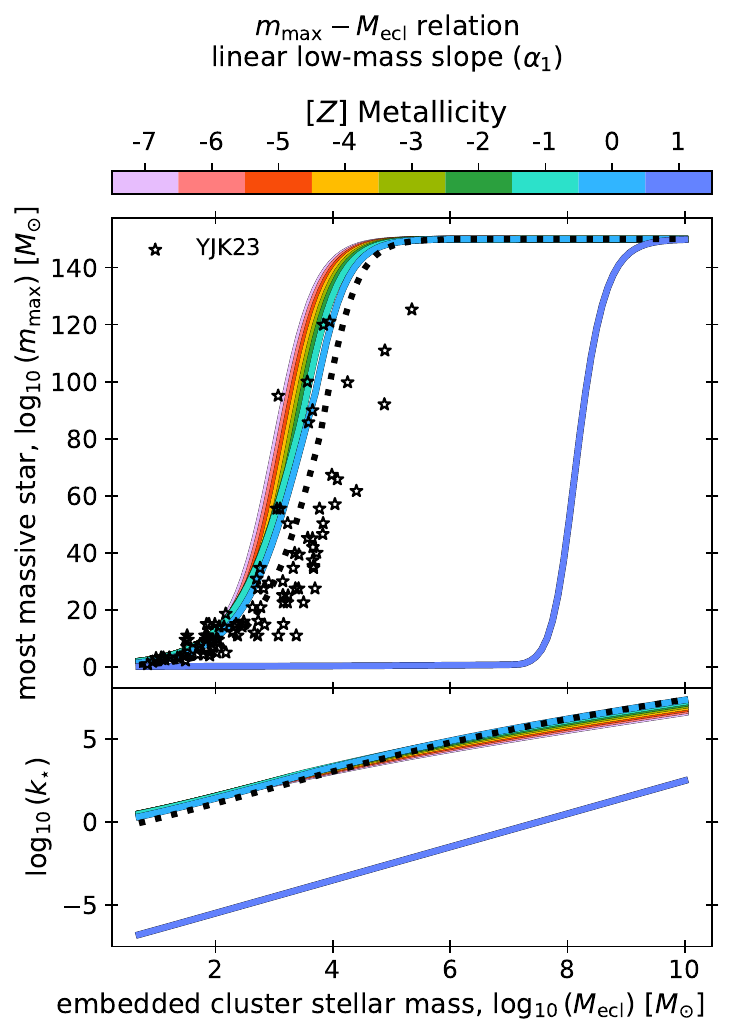}    \includegraphics[width=\columnwidth]{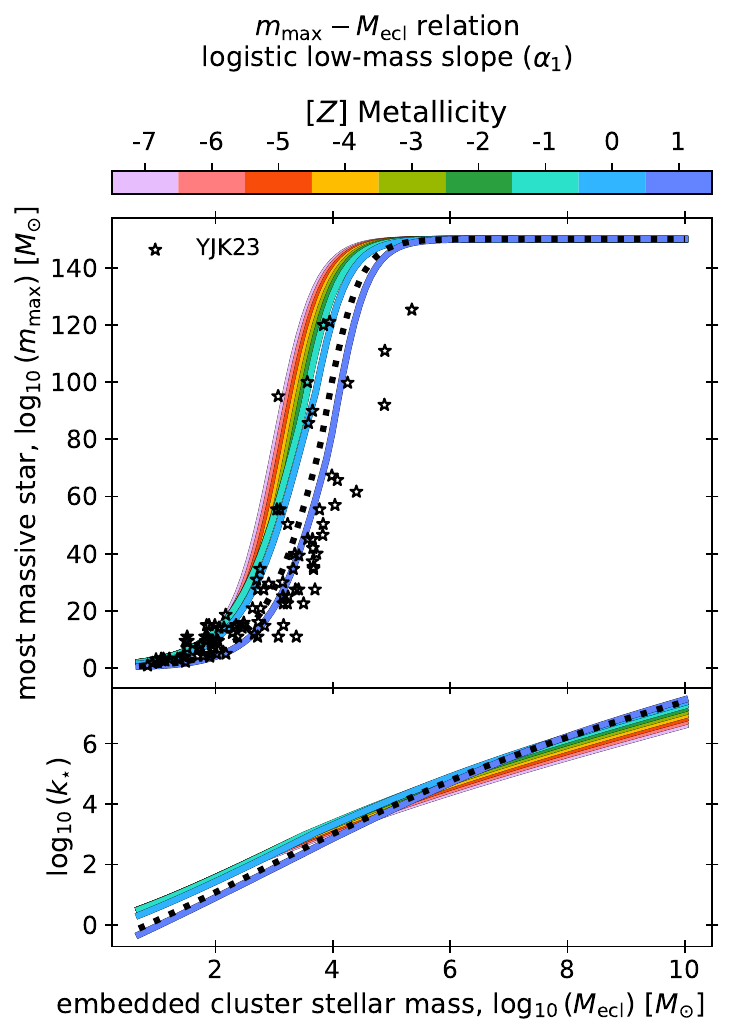} 
    \caption{\emph{Top panels:} The ``$m_{\rm max}$--$M_{\rm ecl}$'' relation, showing the most massive star ($m_{\rm max}$, y-axis) forming within an embedded cluster of stellar mass $M_{\rm ecl}$ (x-axis), as a function of metallicity, [Z] (color-bar). 
    \emph{Bottom panels} Corresponding sIMF normalization constant, $k_{\rm \star}$ (y-axis, Eq.~\ref{eq:sIMF}). 
    \emph{Left panels:} low-mass sIMF slope, $\alpha_1(Z)$, defined with the linear relationship to the metal mass fraction, $Z$ (Eq.~\ref{eq:alpha1}). \emph{Right panels:} $\alpha_1$ defined by the logistic function introduced in this work (Eq.~\ref{eq:alpha1_alternative}). The $m_{\rm max}$--$M_{\rm ecl}$ data (\emph{black stars}) were compiled in \citet{yan+23}, their Table A.1 and constitute less than 5\,Myr old embedded clusters. The \emph{black dotted line} shows the $Z = 2 Z_{\odot} \approx 0.03$, i.e. $[Z] \approx 0.3$ case. }
    \label{fig:mmaxMecl}
\end{figure*}

\myfig \ref{fig:supersolar_sIMF} represents the same sIMFs from \myfig \ref{fig:sIMFsbySFR_logistic}, but allows a direct comparison between the linear and the logistic formulation at supersolar metallicities. Up until $[Z]=0.3$, the two formulations are equivalent. In fact, the top and bottom left panels of Fig.~\ref{fig:sIMFsbySFR_logistic} also apply to the linear case. At $[Z] = 0.5$, which is approximately $Z=3 Z_{\odot}$, the two formulations begin to deviate. 
In this plot it is more clearly evident that an increase in metallicity causes the sIMF to become bottom-heavy. For the lightest-to-intermediate embedded cluster masses, the low gas mass available in the embedded cluster poses a hard limit on the most massive stellar mass generated by that given stellar population. 

Fig.~\ref{fig:mmaxMecl} depicts the ``$m_{\rm max}$--$M_{\rm ecl}$'' relation as a function of metallicity for the linear and logistic $\alpha_1(Z)$ prescriptions. The data points were compiled by \citet{yan+23}, their Table~A.1, with the following literature sources: \citet{km2011, Stephens+2017, WKP13}.

Data deviations from the nominal Solar-metallicity ``$m_{\rm max}$--$M_{\rm ecl}$'' relation as computed here are affected by  the metallicity of the embedded clusters, mergers and ejections of massive stars and stellar-feedback self-regulation of the star-formation process as discussed in detail by \cite{yan+23} and \cite{zhou+2024}. In particular, the frequency of mergers in and ejections from an embedded cluster are subject to the largely unknown properties of massive multiple systems at birth. In some cases, embedded clusters with masses of a few~$10^3 M_\odot$ eject all of their massive stars \citep{ohkroupa2012, ohkroupa2016, ohkroupa2018,  kroupa2025}.

\begin{figure*}[ht]
    \centering
\includegraphics[width=.95\columnwidth]{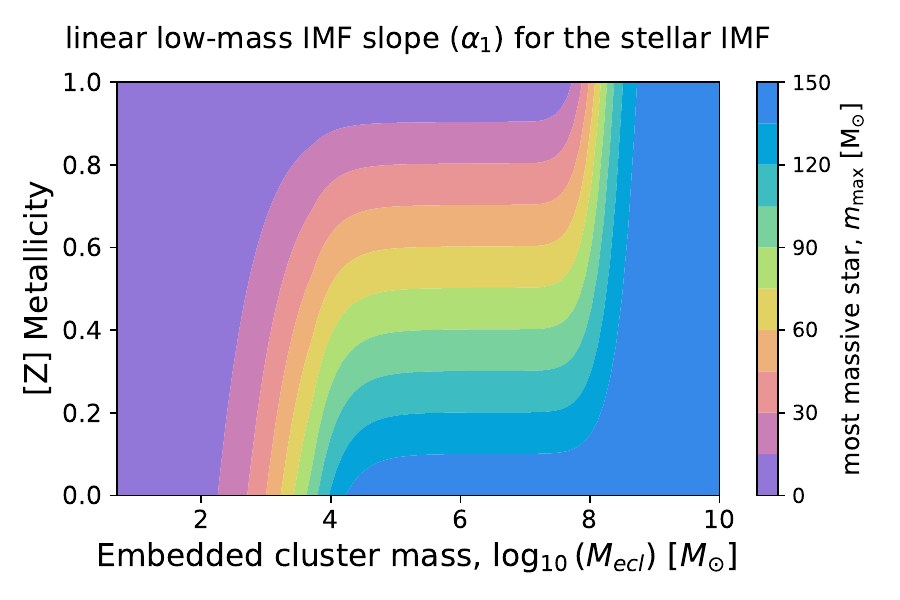}
    \includegraphics[width=.95\columnwidth]{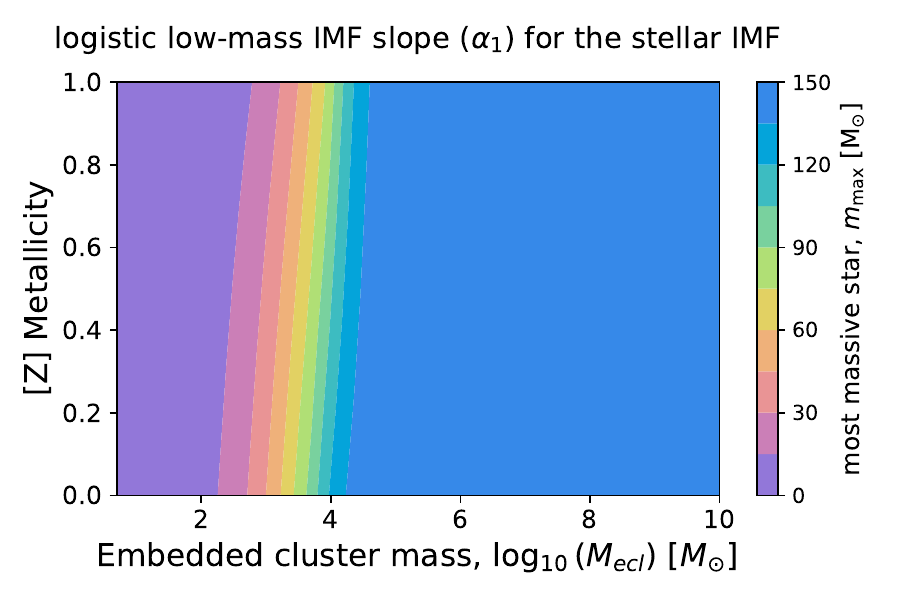}
    \caption{Parameter space of supersolar metallicity ($0<[Z]<1$) and embedded cluster stellar mass ($M_{\rm ecl}$) for the sIMF. The \emph{color-bar} shows the upper limit to the most massive star, $m_{\rm max}$. \emph{Left panel}: Adopts the linear low-mass $\alpha_1$ sIMF slope (Eq.~\ref{eq:alpha1}). \emph{Right panel}: Adopts the logistic $\alpha_1$ slope (Eq.~\ref{eq:alpha1_alternative}). The full metallicity space appeared in \citetalias[][their Fig.~8]{kroupa+2024}.
    }
    \label{fig:ZMecl_mmax}
\end{figure*}
In Fig.~\ref{fig:ZMecl_mmax} we show contours of the most massive stellar masses across the parameter space of embedded clusters and supersolar metallicities. In these panels it becomes clear that for supersolar metallicities, the most massive star that an embedded cluster may form is strongly dependent on the embedded cluster stellar mass, $M_{\rm ecl}$, but it is weakly dependent on metallicity in the logistic case.

\begin{figure*}
    \centering
\includegraphics[width=\columnwidth]{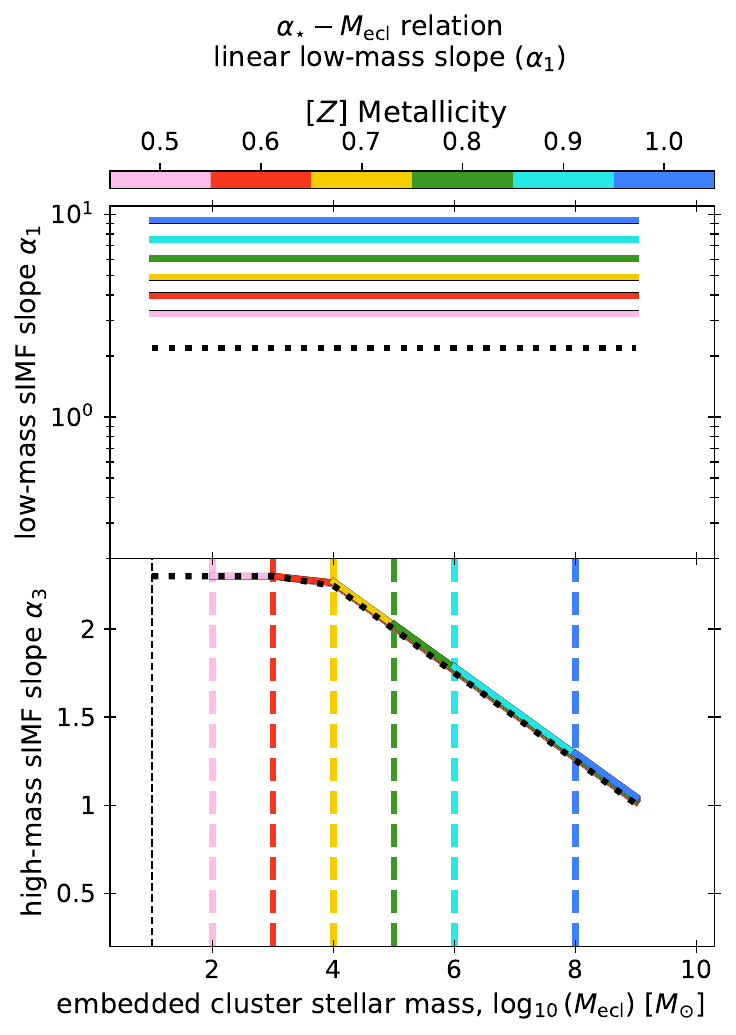}
\includegraphics[width=\columnwidth]{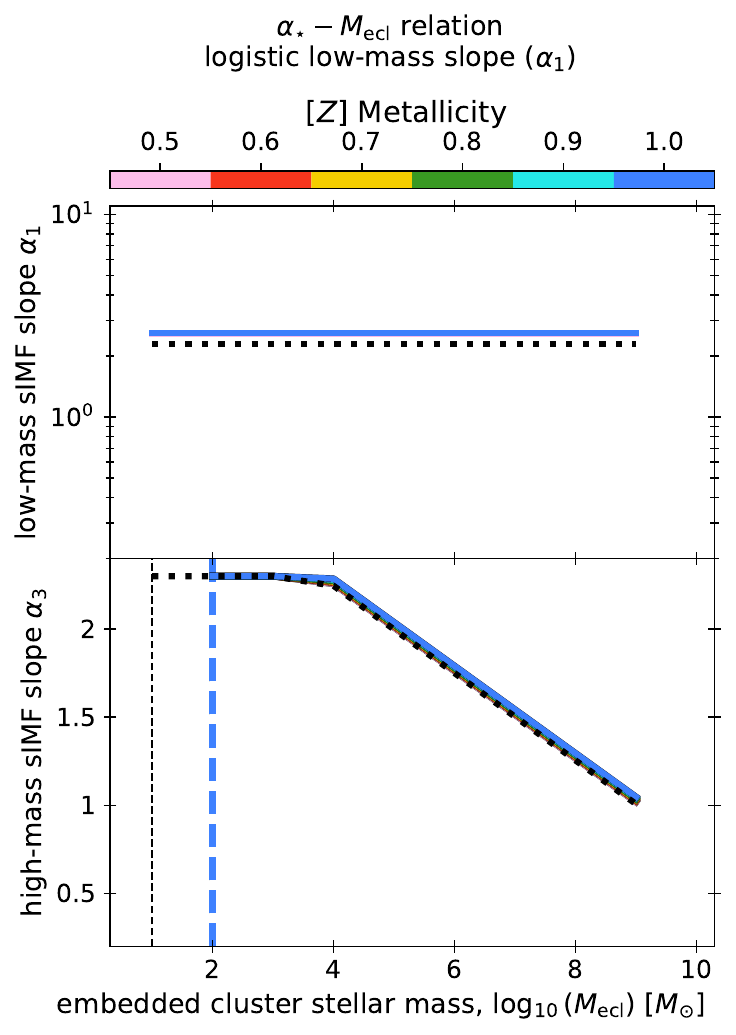} 
    \caption{For the linear prescription (Eq.~\ref{eq:alpha1}, \emph{on the left}) and the logistic prescription of $\alpha_1(Z)$ (Eq.~\ref{eq:alpha1_alternative}, \emph{on the right}), the $\alpha_1 - M_{\rm ecl}$~relation (\emph{top panels}) and $\alpha_3 - M_{\rm ecl}$~relation (\emph{bottom panels}) as a function of metallicity (\emph{color-bar}). 
    The \emph{vertical lines} identify, for each color-coded metallicity, the lowest $M_{\rm ecl}$ where stars above one solar mass ($1 \, M_{\odot}$) begin to be formed. 
    In both plots, the \emph{dotted black line} refers to $Z = 2 Z_{\odot} \approx 0.03$, i.e. $[Z] \approx 0.3$. 
    Up until this metallicity, the two $\alpha_1(Z)$ prescriptions are comparable.  
    While for the logistic prescription, most embedded clusters will produce stars above $1 \, \text{M}_{\odot}$, the linear prescription requires progressively more massive $M_{\rm ecl}$  to reach the solar mass threshold. For example, for $[Z]> 0.6$, $M_{\rm ecl} > 100 \, \text{M}_{\odot}$, and for $[Z] > 1$, $M_{\rm ecl} > 10^7 \, \text{M}_{\odot}$. Furthermore, $\alpha_3$ varies only as long as $m_{\rm max}$ approaches the theoretical upper limit $m_{\rm max *}\approx 150 \, \text{M}_{\odot}$. For metallicities  $[Z]<-2$, $\alpha_1$ effectively remains constant for both $\alpha_1(Z)$ prescriptions. This plot for the full metallicity range is in the Appendix (Fig.~\ref{fig:alpha3alpha1_full}).}
    \label{fig:alpha3alpha1}
\end{figure*}

Fig.~\ref{fig:alpha3alpha1}
shows the sIMF variation described by  Eq.~\ref{eq:alpha1}--\ref{eq:x_rhocl}, namely that at solar metallicity the sIMF remains canonical in molecular cloud clumps that spawn embedded clusters with masses such that $m_{\rm max}\le m_{\rm max*}\approx 150\,M_\odot$. 
Only when $M_{\rm ecl} \gtrsim 10^4 \,M_\odot$, i.e.  when $m_{\rm max} \rightarrow m_{\rm max*}$ (Fig.~\ref{fig:mmaxMecl}), does the sIMF become top-heavy. This can be interpreted in terms of the fragmentation into proto-stellar cores proceeding in a ``normal/canonical'' process up until the clump mass becomes so large that the density ($>10^4\,M_\odot/\text{pc}^3$, or $> 2 \times 10^5\,\text{cm}^{-3}$ in H$_2$, \citealt{McKeeOstriker2007, BerginTafalla2007}) leads to the coalescence of the cores before they can collapse to proto stars, implying a significantly top-heavy sIMF \citep{kroupa+2024}. We note for the first time that the sIMF becomes top-heavy ($\alpha_3 < 2.3$) only when the most massive star in the cluster approaches the theoretical upper limit, i.e., $m_{\rm max} \approx m_{\rm max*}$.

\subsection{The IGIMF theory approach to the gwIMF}
\label{sec:thegwIMF}

\begin{figure*}
            \centering
        \includegraphics[width=\textwidth]{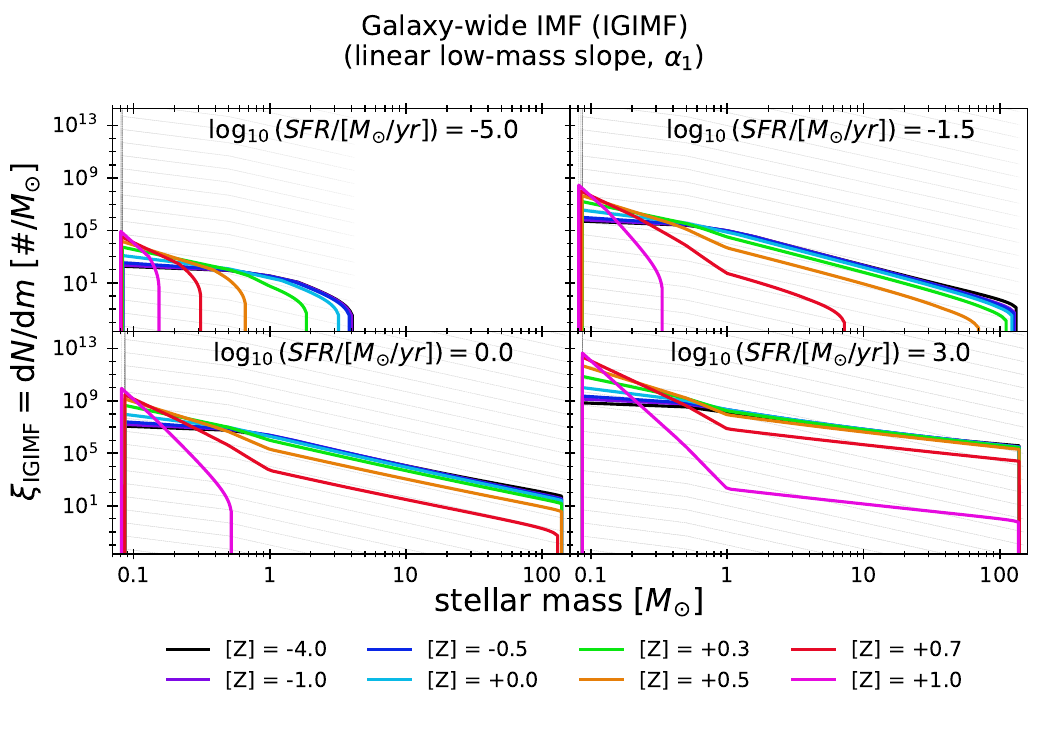}
    \caption{Similar to Fig.~\ref{fig:sIMFsbySFR_logistic},  but for the gwIMF. Here, each panel shows a different SFR, while each line is a different metallicity. IGIMF variations sliced along different SFR values, assuming a linear dependence of $\alpha_1$ on $Z$ (Eq.~\ref{eq:alpha1}). The faint gray background lines show the canonical IMF (Eq.~\ref{eq:canonicalIMF}). Note how galaxies with SFR~$> 1\,M_{\odot}/$yr lack massive stars if they have super-solar metallicity, in contrast to the cases shown in Fig.~\ref{fig:IGIMFsbyZ_logistic}.
    }
    \label{fig:IGIMFsbyZ_linear}
\end{figure*}
\begin{figure*}
        \centering
        \includegraphics[width=0.9\textwidth]{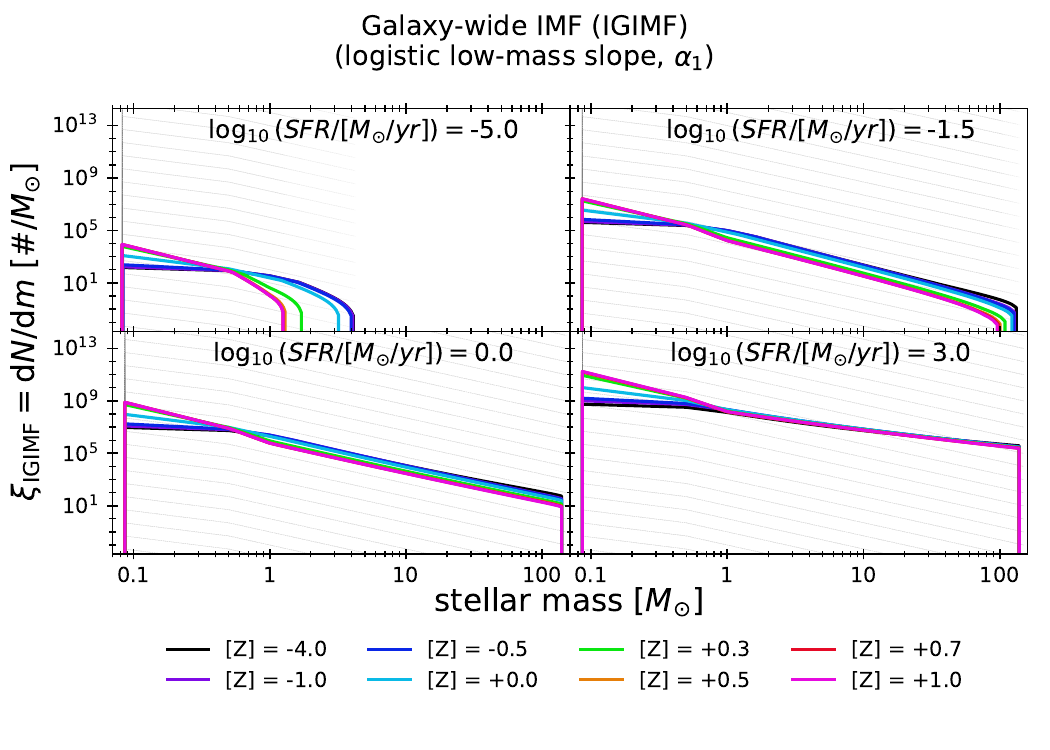}
    \caption{Similar to Fig.~\ref{fig:IGIMFsbyZ_linear},  but for a logistic dependence of $\alpha_1$ on $Z$ (Eq.~\ref{eq:alpha1_alternative}). }
\label{fig:IGIMFsbyZ_logistic}
\end{figure*}

Fig.~\ref{fig:IGIMFsbyZ_linear} and Fig.~\ref{fig:IGIMFsbyZ_logistic} display the gwIMF as described by the IGIMF theory for the linear and logistic $\alpha_1(Z)$ prescription, respectively. In these figures, each panel corresponds to a different galaxy-wide SFR, averaged over the $\delta t$ time interval. The various lines span a wide range of metallicities, with a focus on the supersolar range. In the logistic case (Fig.~\ref{fig:IGIMFsbyZ_logistic}), the gwIMF becomes progressively top-heavier with increasing SFR. The increasing metallicity predominantly affects the low-mass end of the sIMF, which becomes progressively bottom-heavy with increasing metallicity, with only a secondary decrease in massive star production for the highest metallicities. This is a direct consequence of the bottom-heaviness of these gwIMFs: a higher fraction of stellar mass is locked into low-mass stars due to enhanced fragmentation. However, for smaller SFRs, the impact of increasing metallicity becomes progressively more significant. Galaxies with the smallest SFRs will be unable to produce any massive star resulting in core-collapse supernovae ($m \gtrsim 13 \text{M}_{\odot}$, \citealt{Limongi2017}), regardless of the $\alpha_1(Z)$ prescription. In the case of the linear prescription (Fig.~\ref{fig:IGIMFsbyZ_linear}), however, the bottom-heaviness induced by metallicity-driven fragmentation  becomes so pronounced that it affects the ability to produce massive stars. Interestingly, for low to average SFRs ($\lesssim 1 \, \text{M}_{\odot}$/yr), metallicities above $[Z]> 0.5$ begin to dramatically reduce star formation of $\approx 1 \, \text{M}_{\odot}$ stars.

\begin{figure*}[ht]
    \centering
\includegraphics[width=\columnwidth]{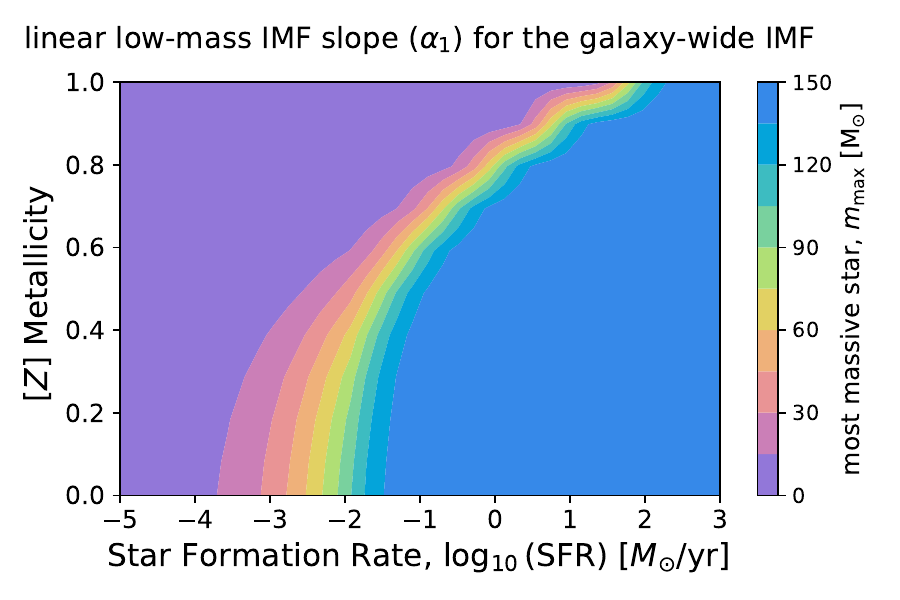}
\includegraphics[width=\columnwidth]{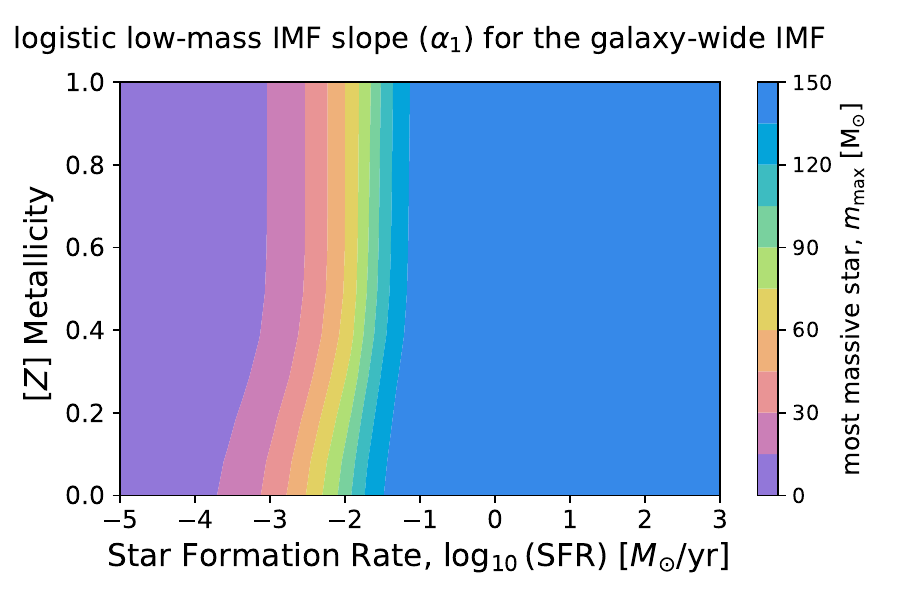}
    \caption{Similarly to Fig.~\ref{fig:ZMecl_mmax}, parameter space of supersolar metallicity ($0<[Z]<1$) and galaxy-wide SFR, averaged over timesteps $\delta t = 10^7$~yr, for the gwIMF. The \emph{color-bar} shows the upper limit to the most massive star, $m_{\rm max}$. \emph{Left panel}: Adopts the linear low-mass $\alpha_1$ sIMF slope (Eq.~\ref{eq:alpha1}). The IGIMF depends on the metal~mass~fraction. The undulations in the contours arise from the logarithmic dependence of metallicity, $[Z]$, on the metal~mass~fraction, $Z$ (see Sec.~\ref{sec:metallicity}). \emph{Right panel}: Adopts the logistic $\alpha_1$ slope (Eq.~\ref{eq:alpha1_alternative}). Note that the $M_{\rm ecl}$ in Fig.~\ref{fig:mmaxMecl} was computed on a finer grid than the SFR in the current plot.}
    \label{fig:ZSFR_mmax}
\end{figure*}

Fig.~\ref{fig:ZSFR_mmax} for the gwIMF, similarly to Fig.~\ref{fig:ZMecl_mmax} for the sIMF, shows the mass of the most massive star as a function of its parameter space. In this case, this is the metallicity and SFR.  In the logistic $\alpha_1(Z)$ prescription, the value of the most massive star that may form in a galaxy is fairly independent of metallicity. Meanwhile in the linear case, for metallicities $[Z] \gtrsim 0.5$, the formation of massive stars is steeply constrained even at fairly high metallicities. 

The IGIMF theory generically predicts gas-rich dwarf galaxies with a true $\bar{\psi} < 10^{-5}\,M_\odot$/yr to have H$\alpha$-invisible star formation because only stars with masses $m \le m_{\rm max}<1\,M_\odot$ can form (Fig.~\ref{fig:ZSFR_mmax}). The existence of dwarf galaxies with H$\alpha$-dark SFR has already been noted by \citep{pawk2007}. These authors also point out that dwarf galaxies have, due to their top-light gwIMFs, significantly larger true SFRs than the estimates via the standard H$\alpha$ luminosity implies (see also \citealt{jerabkova+2018, haslbauer+2024}), with important repercussions for the cosmological matter cycle.  
But as is also evident from Fig.~\ref{fig:ZSFR_mmax}, if the linear low-mass IMF slope is the correct description then H$\alpha$-invisible star formation will also occur when $[Z]>0.6$. This means we would have massive disk galaxies with a large amount of cooling supersolar-metallicity gas with $\bar{\psi} > 1\,M_\odot$/yr but no detectable star formation as only stars with $m \le m_{\rm max}<1\,M_\odot$ form. Such galaxies, if they exist, would have an extremely slowed chemical enrichment because only AGB stars and stellar mergers \citep{wang+2020} would be contributing new metals the inter stellar medium.

\subsection{A special case of the IGIMF theory at high redshifts}
\label{sec:concordanceIMF}

\begin{figure}
    \centering
    \includegraphics[width=\columnwidth]{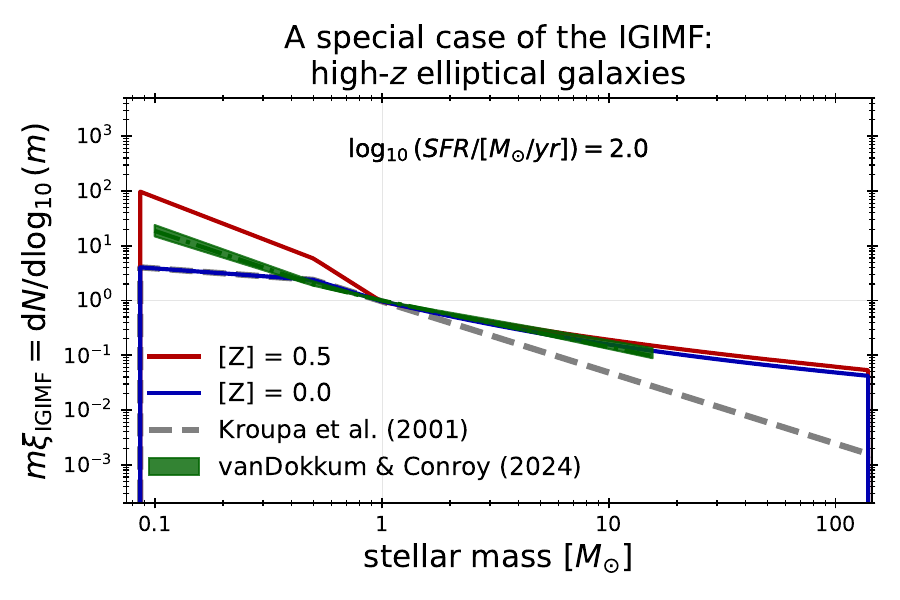}
    \caption{Elliptical galaxies deviate significantly from the canonical IMF (see Sec.~\ref{sec:canonicalIMF}). From recent JWST observations of high-redshift ($z >10$) elliptical galaxies, \citet{vandokkumconroy2024} derived a top-heavy and bottom-heavy IMF they refer to as ``concordance IMF''. It is a special case of the IGIMF, here plotted in the logistic formulation (see Sec.~\ref{sec:mass-weighted}). It corresponds to  starburst (SFR~$\approx 100 \, \text{M}_{\odot}/\text{yr}$) galaxies enriched to solar (\emph{blue line}) and supersolar abundances (\emph{red line}, $[Z]\approx0.5$, $Z \approx 3 Z_{\odot} \approx 0.04$).
    }
    \label{fig:concordanceIMF}
\end{figure}

In Fig.~\ref{fig:concordanceIMF} we show that the IMF proposed in \cite{vandokkumconroy2024} is a special case of the IGIMF.
The formulation by \citealt[][and references therein, ]{vandokkumconroy2024} constructs an average gwIMF (referred by them as the ``Concordance IMF'') 
for elliptical galaxies that are bottom-heavy. 
Above $1\,M_{\odot}$, they adopt a top-heavy gwIMF to reproduce the small mass-to-light ratios inferred for $z>10$ galaxies observed with JWST. Such low ratios are required in the standard $\Lambda$CDM framework for young, low-mass galaxies to match the observed luminosities at these redshifts (but see \citealt{haslbauer+2022b} for a discussion).

In order to link the high-redshift data with the present-day elliptical galaxies and star-forming disk galaxies they suggest that $\alpha_{1,2,3}$ are functions of the present-day velocity dispersion of a galaxies \citep[][their Eq.~6]{haslbauer+2022b}. 

The \citet{vandokkumconroy2024} approach thus captures aspects that are contained in the IGIMF theory. To demonstrate this, their gwIMF formulation is rewritten in terms of IGIMF solutions that match their solution shown in \citet{vandokkumconroy2024}, their Fig.~3. We achieve this agreement through specific combinations between the metallicity $Z$ and SFR, as shown in Fig.~\ref{fig:concordanceIMF}. Namely, the ``bottom-heavy'' portion occurs at supersolar metallicities, while the ``top-heavy'' portion is due to high SFR.

The IGIMF theory provides physical motivations to the variability of the gwIMF,
and can be calculated for any combination of these parameters $\bar{\psi}_{\delta t}, Z$, allowing also self-consistent chemical enrichment histories to be computed for ultra-faint dwarf galaxies to massive ellipticals starting from their formation times \citep{yan2019, yan20, yan+21} together with photometry and spectra that match observed galaxies \citep{haslbauer+2024, zonoozi+2025}. 
The IGIMF theory provides a self-consistent framework for computing the stellar population evolution in any galaxy, including a formulation for how young, low-dispersion galaxies evolve into the massive systems observed today.
Meanwhile, the ``Concordance IMF'' is tailored to elliptical galaxies, with extensions to other systems parameterized only by their present-day velocity dispersion, without a clear connection to the evolution of a galaxy.

Given the above approaches to the gwIMF, it is noteworthy that all have in common that it needs to be top-heavy at high redshift and bottom-heavy in deep potential wells. From the propositions, only the IGIMF explicitly links the physical properties of molecular cloud clumps to the galaxy-wide stellar population within a self-consistent computable theory. 

\section{Discussion} \label{sec:discussion}

In this section we first briefly overview other approaches to a variable gwIMF. We then discuss three aspects of the IGIMF:  its conceptual grounding, its empirical evidence, and the concept of optimal sampling.

\subsection{Other approaches to the gwIMF}
\label{sec:othergwIMFs}

The IGIMF approach is distinct from other gwIMF approaches that do not build the galaxy-wide stellar population based on the star's formation sites but invoke instead galaxy-wide physical properties. 

The formulation by \cite{jermyn18, steinhardt22a, steinhardt22b}
assumes the form of the gwIMF to be invariantly canonical but that the masses at which the power-law indices $\alpha_{1,2,3}$ shift to different stellar masses depend on the temperature of the gas in the galaxy which depends also on the redshift. As discussed in \cite{kroupa+2024} this approach has the drawback that the temperature is not readily accessible once the stars have formed such that reconstructing the sIMFs from a gwIMF becomes difficult. This approach thus does not link the galaxy-wide IMF to the physics of star-forming sites in different locations in a galaxy. The IGIMF in contrast constructs a gwIMF through the integration at a given time of all the stellar masses born in embedded clusters across the mass distribution of the very same embedded clusters, and it contains an implicit temperature dependency through the distribution of molecular cloud clump densities that enter the SFR of the galaxy.

\subsection{Stars are born in embedded clusters within molecular clouds} \label{sec:intro-foundation}

The occurrence of star formation within embedded clusters has been confirmed by Galactic surveys \citep{carpenter+1995, carpenter+2000, porras+2003, LadaLada2003, megeath+2016, winston+2020}, by Galactic binary star properties \citep{kroupa1995a, kroupa1995b, mk2011, dabringhausen+2022, kroupa2025}, and extra-galactic observations \citep[][and references therein]{dinnbier+2022, cook+23}. 
However, the fact that star formation occurs primarily within embedded clusters is non-trivial. In fact, many young stars are observed outside of embedded clusters \citep[e.g.,][]{gieles+12}, leading some to believe that stars may form elsewhere within the molecular clouds. However, such young stars were likely not born at their current location. Several authors \citep{OK2015, dinnbier+2022} have shown that the rapid early evolution of the embedded cluster disperses young stars widely throughout the star-forming region. This process occurs as quickly as one Myr since their birth \citep{OK2016}. 

\begin{figure}
    \centering
    \includegraphics[width=\columnwidth]{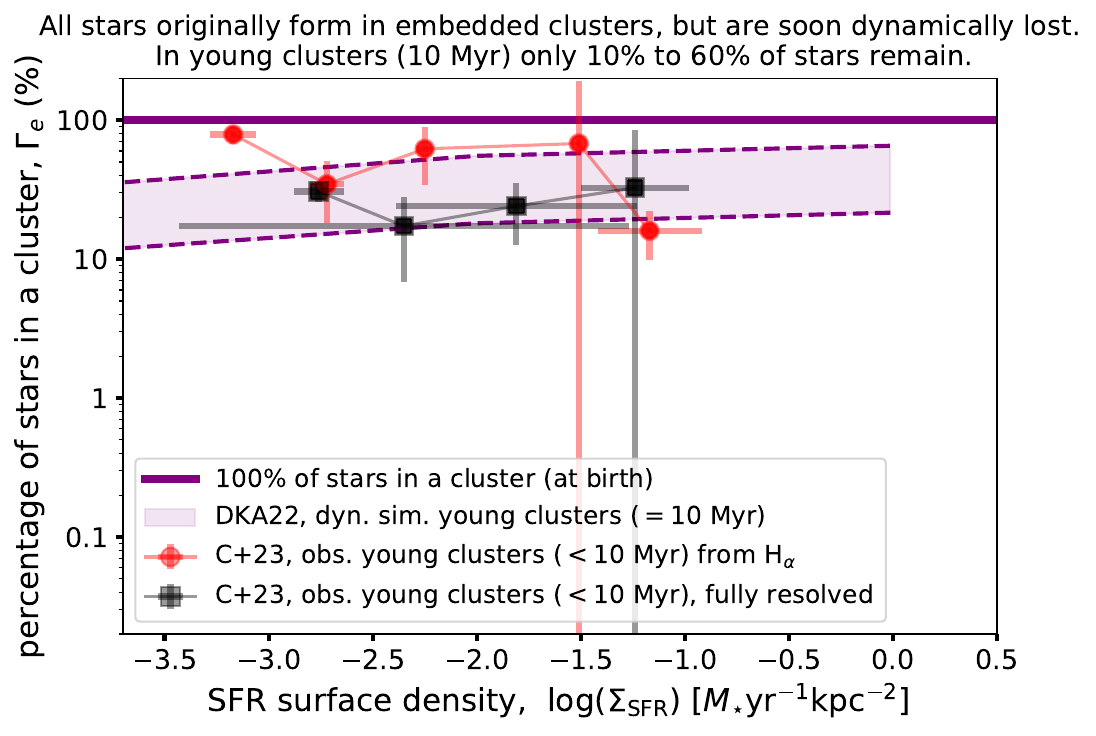}
    \caption{Percentage of stars ($\Gamma_e$) that remain in post-gas expulsion clusters 10~Myr after cluster formation, as a function of the embedded cluster SFR surface density, $\Sigma_{\rm SFR}$.
    At birth, we assume that all stars form within embedded clusters (i.e., no stars form in isolation). This is the foundational assumption of the IGIMF theory. 
    The \emph{solid~purple~line} at $\Gamma_e = 100$\% represents this initial condition, which is the condition implemented in N-body simulations 
    \citep[][DKA in the figure]{dinnbier+2022}, which also take into account gas expulsion and a star-formation efficiency of 33\%.
    These simulations predict that after 10~Myr, only 10–60\% of the stars originally formed in embedded clusters remain bound 
    (\emph{shaded~purple~region}). 
    These theoretical results are consistent with observations, where the binned data for young ($<10$~Myr) clusters is taken from \citet{cook+23}, 
    their Figure~12.}
    \label{fig:Cook23}
\end{figure}

Fig.~\ref{fig:Cook23} contains simulation results, represented by the purple lines and shaded regions,  as well as observations from the LEGUS survey on dwarf~galaxies, represented by the red and black bins \citep{cook+23}. {Black~bins} represent data for resolved stellar populations (where individual stars are detected), while {red~bins} correspond to results for unresolved populations (where the SFR is inferred from integrated light). All young resolved clusters fall within the limits predicted by dynamical evolution models, while unresolved populations preferentially reside above the conservative simulation results. The axis limits were chosen to match those in Figure 12 of \citet{cook+23}. Both observations and simulations consider young ($\leq 10$~Myr) embedded clusters. The purple solid line is the IGIMF theory rationale: all stars are born in embedded clusters. Dynamical N-body evolution reduces the fraction of stars remaining within the embedded clusters (purple shaded area) to values consistent with observational uncertainties for similarly young clusters, namely, between 40\% and 90\% of all stars are dynamically lost from the cluster within the short time-frame of 10~Myr \citep{dinnbier+2022}.
Aside from typical molecular clouds, embedded clusters can also form in filamentary distributions along probable shock fronts, provided that the required density and temperature conditions are satisfied. Possible examples are the 100~to~300~pc-long relic filaments discovered by \citet{Jerabkova+2019, Beccari+2020}.

\subsection{Extragalactic evidence that star formation occurs in embedded clusters}

There is compelling evidence that stars form in localized discrete regions, i.e.,  in the regions of molecular clouds that undergo gravitational collapse \citep[][their Sec.~3.1]{kroupa+2024}. Such regions must, by necessity for collapse (to form stars) be very compact, since they need to achieve a sufficient density for the collapse to ensue and to form stars. During the collapse, always more than one star forms, since these regions contain much more mass than that which ends up in a single low-mass star (star formation being inefficient, less than some 30\% of the gas ending up in stars, \citealt{BattistiHeyer2014}).  These regions thus must form embedded clusters that may contain just a few multiple systems to many millions of binaries at birth (the latter in interacting galaxies, \citealt{mahani+21}). 
At the low-mass end, good examples are the small embedded clusters in the Taurus–Auriga region \citep[e.g. the ``nests'' of][]{joncour+18} and in $\rho$-Oph \citep{km2011}. Similarly, Spitzer observations of the Orion molecular clouds reveal that these clusters have low star formation efficiencies (\citealt{megeath+2016}, their Fig.~25) and that the youngest pre-main sequence stars are strongly clustered (\citealt{megeath+2016}, their Fig.~26).

Dynamical population synthesis simulations support the observational evidence, and demonstrates that the binary star properties observed in star-forming regions naturally lead to the observed Galactic-field binary star population \citep{kroupa1995a, kroupa1995b}.
\citet{kroupa2005} advanced the hypothesis that embedded star clusters can be treated as the fundamental building blocks of galaxies.

Embedded clusters are dynamically very active and so they expel stars quite rapidly (in less than a Myr, \citealt{ohkroupa2012, dinnbier+2022}) due to the high birth multiplicity fraction. In fact, a binary can contain much binding energy that can be transformed into kinetic energy when an encounter with another star or binary occurs in the embedded cluster, e.g., \citealt{kroupa2025}. The embedded clusters expel most of their mass as gas through the feedback from the forming stars therewith they expand - \citet[e.g.][]{dinnbier+20}. 
Embedded clusters do not typically interact while forming, since molecular clouds have long free fall times such that the embedded cluster form and expel their residual gas well before merging \citep{Kroupa1998a, Kroupa1998b}, while their merging becomes relevant only in massive star~cluster complexes \citep{mahani+21}.

Any gwIMF must be consistent with galaxy-scale properties as well as the observed sIMFs in molecular clouds. The IGIMF does just that: by considering the resolved and dynamically well-modeled star clusters and ultra-compact dwarf galaxies, it gauges the dependency of the sIMF from the physical attributes of embedded clusters \citep[][their Sec.~4.3]{kroupa+2024}. What emerges is a theory that accounts for many galaxy properties.

\subsection{The sIMF must be variable} \label{sec:intro-variation}

Metal-rich gas cools significantly faster than low-metallicity gas due to a larger number of allowed electron transitions \citep{PloeckingerSchaye2020}. 
From a theoretical standpoint, inefficient cooling at low metallicity reduces gas fragmentation, resulting in fewer low-mass stars compared to higher-metallicity environments. Consequently, the sIMF at low metallicities is expected to be deficient in low-mass stars relative to the canonical form (for the definition of the \emph{canonical sIMF}, see Sec.~\ref{sec:canonicalIMF}). Additional physical parameters that are expected to affect the shape of the sIMF are cosmic-ray heating of the clump \citep{papadopoulos2010}, magnetic fields and rotational shear which would inhibit the formation of low-mass stars near a super-mass black hole (for reviews see \citealt{kroupa+2013, hennebellegrudic2024, kroupa+2024}) and especially the coalescence of massive cores in the dense innermost regions of massive clumps,
before said cores can collapse to a proto-star \citep{dib+2007}.

Concerning observational evidence, in star-burst conditions, embedded clusters may range in mass up to a few hundred thousand to many hundreds of millions of stars \citep{BK2018}, with radii of $< 1$~pc \citep{MarksKroupa2012}. On the opposite end, the smallest embedded clusters generate a total of $5\,M_\odot$ in newly formed low-mass stars (\citealt{km2011, joncour+18}). 
Using binary stars as the tracers of the maximum possible density an open cluster could have had at birth, \cite{MarksKroupa2012} found the birth half-mass radii of the embedded clusters to follow the relation $r_{\rm h} \approx 0.1 \times \left(M_{\rm ecl}/M_\odot\right)^{0.13}$ (\citealt{MarksKroupa2012}, their Eq.~7).

Interestingly, \citet{hsu+2012} found that even moderately-massive embedded clusters which generate $\approx 50-100\,\text{M}_{\odot}$ in new stars do not generate massive stars. These observations indicate that there is immense variability in  sIMF. Consequently, the gwIMF will tightly depend on the mass distribution of the embedded clusters. Molecular clouds that form low-mass to moderately-massive embedded clusters may have a systematically different sIMF than molecular clouds generating more massive embedded clusters. 
\citet{weidner+2006} deduced that the most massive star in an embedded cluster, $m_{\rm max}$, appears to be strongly related to the stellar mass of the embedded cluster itself, $M_{\rm ecl}$. This relation is further investigated and compared to other scenarios in \citet{yan+23}.

On top of being dependent on $M_{\rm ecl}$, the sIMF was shown to further depend on the molecular~cloud clump metallicity, $[Z]$, as well as its density, $\rho_{\rm cl}$. The gas density $\rho_{\rm cl}$ eventually generates a total stellar mass, $M_{\rm ecl}$.  A first hint of this variability was advanced in \citet{kroupa2002}.
This metallicity~dependence was robustly confirmed for the first time in the solar neighborhood by \citet{Li+23}.

Detailed modeling of globular star clusters, ultra-compact dwarf galaxies and nearby massive star-burst clusters, when compared to observational data indicate that the sIMF systematically varies with the density and metallicity of the star-forming gas (\citealt{dabringhausen+09, dabringhausen+12, marks+12, wirth+2022} and references therein). The variation is such that at large $\rho_{\rm cl}$ the sIMF becomes top-heavy. For metallicities as low as $[Z] \approx -1$,  the sIMF of moderately massive embedded clusters also becomes top-heavy, and additionally it becomes bottom-light, i.e., the number of massive stars is enhanced while the number of low-mass stars is suppressed. At larger $Z$, the opposite is true: the sIMF becomes bottom-heavy and for sufficiently low-mass embedded clusters, top-light, i.e., the number of massive stars is suppressed while the number of low-mass stars is enhanced.

\subsection{The sIMF is inconsistent with stochastic sampling}\label{sec:intro-stochastic}

If fragmentation of a gas clump depends on the metallicity, then the shape of the sIMF should remain invariant at fixed $Z$. Since low-mass clumps can only form a small number of stars, the mass of the most-massive star ($m_{\rm max}$) is expected to depend on the total stellar mass ($M_{\rm ecl}$) of the embedded cluster in which it forms. Such a relation ($m_{\rm max}-M_{\rm ecl}$) has indeed been found at near-Solar metallicity \citep{weidner+2006, weidner+2010, weidner13, yan+23}. Given the observational uncertainties, the distribution of $m_{\rm max}$ values at a given $M_{\rm ecl}$ informs us of the possible stochastic, i.e. probabilistic nature of the sIMF.

The sIMF has been shown to vary both with metallicity  as well as with clump mass (Sec.~\ref{sec:intro-variation}), and it is inconsistent with a stochastic origin:
by investigating the dispersion of  $m_{\rm max}$ as a function of $M_{\rm ecl}$, \citet{yan+23} found that this dispersion is too small to be consistent with the sIMF being a stochastic probability distribution function. 

Similarly to the above, \citet{xu2024,Jiao2025baobab} identified a relation between the most massive core and the mass of the clump in which it forms. Regarding the core mass distribution within a clump, optimal sampling (see Sec.~\ref{sec:intro-optimalS}) is favored over stochastic sampling.

Additionally for solar to moderately sub-solar metallicities, the sIMF of massive stars is characterized by a power-law index, $\alpha_3$, for stars exceeding one solar mass (1 $\text{M}_{\odot}$). The various estimates of $\alpha_3$ are found to be too close to the Salpeter slope (Eq.~\ref{eq:salpeter}, see also \citealt{kroupa2002}, their Fig.~6), and therefore $\alpha_{\rm S55}$. Stochastic sampling is unable to generate the small observed standard deviation \citep{kroupa2002, yan+23}.
\citet{kroupa+2013} proposed that the sIMF might be an optimally sampled distribution function (ODF). Drawing stars from an ODF leads to a distribution without Poisson scatter and can be interpreted physically to mean that the formation of an embedded cluster in a molecular cloud core is highly self-regulated.

\subsection{Optimal sampling}\label{sec:intro-optimalS}

\citet{kroupa+2013} introduced the process of optimal sampling in the context of selecting  
stars from the sIMF, leading to the equations  shown in Eq.~\ref{eq:stellar_optimalsampling}. 
The principle behind optimal sampling is straightforward: 
given a shape of the sIMF, the quantity $M_{\rm ecl}$ needs to be distributed over an ensemble of stars so that this ensemble obeys the sIMF exactly, without a Poisson dispersion.
Here the empirically discovered most-massive star--embedded-cluster-mass relation (see Fig.~\ref{fig:mmaxMecl} below) was instrumental as it implies there to be one well-defined most-massive star in an embedded cluster of stellar mass $M_{\rm ecl}$.
The mathematical treatment of optimal sampling was improved by \cite{schulz15}.

Under this premise, two physically identical gas clumps will thus produce the exact same sequence of stellar masses. As a consequence,
there exists only one most massive star of mass $m_{\rm max}$ in every single stellar population of mass $M_{\rm ecl}$, and the sIMF shape is obeyed exactly (Eq.~\ref{eq:stellar_optimalsampling}).

As discussed in \cite{kroupa+2024}, their Sec.~3.3, the existence of an upper limit to stars can broadly be understood theoretically through an accreting protostar feedback becoming too strong for further gas to accrete on to it, but the details depend on the geometry (accretion from a disk), the metallicity, accretion/mergers with other cores and magnetic fields and remain uncertain \citep{clarke2006, zinneckeryorke2007}. That $m_{\rm max*}\approx 150\,M_\odot$ is an empirical finding from the cut-off of star counts in very young stellar populations that should, purely on statistical grounds, hold more massive stars if they were to exist such as in the star-burst cluster R136 in the sub-Solar metallicity Large Magellanic Cloud \citep{weidnerkroupa2004, koen2006}, the metal-rich starburst cluster Arches near the Galactic center \citep{figer2005}, various very young populations in the Milky Way and the Magellanic Clouds \citep{oeyclarke2005} and the very young cluster Pismis~24 \citep{maizappellaniz2007}.
A few massive stars with masses surpassing $150\,M_\odot$ have indeed been observed but these are most likely mergers \citep{banerjee+2012, ohkroupa2018}, given the constraints on their ages and the persisting number cutoff noted above. While a metallicity dependency of $m_{\rm max*}$ is expected, there is no empirical evidence for this yet.

We suggest that the interpretation of the sIMF as an optimally sampled function rather than a probability density distribution, may be related to the principle of maximum entropy which states that, under known constraints, the least biased distribution is the one that maximizes entropy. It provides a way to determine the most statistically neutral and constraint-respecting stellar mass function without introducing unnecessary assumptions 
(Gjergo, Zhang \& Kroupa, submitted).
In the principle of maximum entropy, a natural cut-off (i.e., $m_{\rm max}$) emerges without invoking randomness, for example, due to the capacity of substructures within the molecular gas clump to collapse and form a massive star. It is a process that ensures mass conservation and avoids arbitrary statistical assumptions.

\section{Conclusions} \label{sec:conclusion}

We investigated two limiting cases for the variation of the sIMF and gwIMF at high metallicities.
For subsolar to moderately supersolar metallicities, the formation of low-mass stars ($< 1 \, M_{\odot}$) is particularly sensitive to the metal mass fraction, $Z$, of the star-forming gas. The slopes describing the low-mass regime of the sIMF, $\alpha_1$ and the dependent $\alpha_2$  (Eq.~\ref{eq:alpha1}), increase linearly with $Z$.  
Higher metallicities enhance fragmentation efficiency, leading to the formation of a larger number of lower-mass stars.
This trend is well established in environments with metallicities up to about three times solar ($[Z] = 0.5$). 
However, star formation is known to occur at even higher metal mass fractions of 7 to 10 times solar (Fig~\ref{fig:alpha1alternative}). The sIMF and gwIMF under these conditions are still poorly constrained.

The two limiting cases are as follows: In the first scenario, \emph{linear $\alpha_1$}, we extrapolate the known linear dependence  of $\alpha_1$ on the metal mass fraction (Eq.~\ref{eq:alpha1}) to the highest metallicities. In the second scenario, 
we assume that fragmentation efficiency has already saturated by $[Z] \approx 0.5$, such that $\alpha_1$ remains constant for higher metallicities. 
To describe the smooth transition from a metallicity-sensitive to a metallicity-insensitive regime, we introduce a logistic $\alpha_1$-$Z$ form, the  \emph{logistic $\alpha_1(Z)$} (Eq.~\ref{eq:alpha1_alternative}).

We find that although the linear $\alpha_1$-$Z$ relation reproduces the distribution of low-mass stars, its extension to extreme supersolar regimes suppresses the formation of massive stars. In this case, gas fragmentation would become too efficient to allow the assembly of high-mass stars. Alternatively, the logistic formulation yields an IMF that remains invariant with increasing metallicity.
We conclude that for metallicities in the range $0.5 < [Z] <1.0$, if fragmentation remains sensitive to increasing metal content, the formation of massive stars may be strongly suppressed.

Under the assumption that the sIMF is an optimally~sampled distribution function (Sec.~\ref{sec:intro-optimalS}) dependent on metallicity and density, we quantified 
the maximum stellar mass, $m_{\rm max}$, that can form in an embedded cluster whose total stellar mass is $M_{\rm ecl}$. This is known as the $m_{\rm max} - M_{\rm ecl}$ relation. We found that it systematically shifts toward higher embedded cluster masses with increasing metallicity.
Our results show that:
\begin{enumerate}
  \renewcommand{\labelenumi}{(\roman{enumi})}
    \item For a given $M_{\rm ecl}$, the formation of massive stars is strongly suppressed under a linear $\alpha_1(Z)$ prescription at $Z \gtrsim 2 Z_{\odot}$, but remains within familiar ranges under the logistic model. 
    \item The gwIMF reflects this suppression and steepens significantly at high metallicities and low SFR.
\end{enumerate}

This work shows that if molecular cloud fragmentation increases with metallicity, as described by the linear $\alpha_1(Z)$ prescription, then at supersolar metallicities ($[Z] \geq 0.5$), even galaxies with high star formation rates  would appear as low-surface-brightness systems (Fig.~\ref{fig:IGIMFsbyZ_linear}), since their star~formation would be dominated by low-mass stars. Such galaxies, if they exist, would continue to accumulate stellar mass without substantial chemical enrichment, apart from contributions by AGB stars and stellar mergers.

In all IGIMF models, dwarf galaxies with very low star formation rates (SFRs~$<\,{\rm few} \times 10^{-5}\,M_\odot\,\text{yr}^{-1}$) do not form massive stars and are therefore undetectable in H$\alpha$. We show however that, under the linear $\alpha_1(Z)$ prescription, even galaxies with high gas-phase metallicities and SFRs exceeding $1\,M_\odot\,\text{yr}^{-1}$ may remain H$\alpha$-dark, because star formation would be limited to low-mass stars only.

In Sec.~\ref{sec:concordanceIMF}, we show that the ``concordance IMF'' \citep{vandokkumconroy2024} is a special case of the IGIMF for supersolar starburst conditions. Furthermore, our results highlight, for the first time, that the stellar IMF becomes significantly top-heavy, with $\alpha_3 < 2.3$, only when the most massive star in the embedded cluster reaches the fundamental upper limit ($m_{\rm max} \approx m_{\rm max*}$).

Assuming that, after gravity, metallicity is the dominant quantity affecting star formation, this work offers a theoretical framework for interpreting IMF variations in metal-rich environments. The impact on galaxy evolution, chemical enrichment, and black hole formation histories is substantial, especially in early-type galaxies and the bulges of spirals. Future high-resolution observations of massive star formation in metal-rich environments will be crucial to constrain the shape of the sIMF and test the predictions made here.
Retrieving the true underlying gwIMF is particularly important when observing high~redshift galaxies, where individual stars cannot be resolved and yet the metallicity may already be highly supersolar. Given the central role of the sIMF in galaxy evolution models, it is critical to constrain the gwIMF through indirect observational probes.

Lastly, we released with this publication a user-friendly Python package, \texttt{pyIGIMF} (Appendix~\ref{sec:appendix-software}), which enables the rapid computation of the gwIMF as a function of SFR and metallicity.

\begin{acknowledgements}
E.G., Z.Y. and Z.Z. acknowledge the support of the National Natural Science Foundation of China (NSFC) under grants NOs. 1251101411, 12173016, 12041305.
E.G., Z.Y. and Z.Z.  acknowledge the science research grants from the China Manned Space Project with NOs. CMS-CSST-2021-A08 (IMF), CMS-CSST-2021-A07.
E.G., Z.Y. and Z.Z. acknowledge the Program for Innovative Talents, Entrepreneur in Jiangsu. 
Z.Y. acknowledges support from National Natural Science Foundation of China under grant number 12203021, the Jiangsu Funding Program for Excellent Postdoctoral Talent under grant number 2022ZB54, the Fundamental Research Funds for the Central Universities under grant number 0201/14380049. P.K. acknowledges support through the DAAD-Eastern-European Exchange programme between Bonn and Prague.
\end{acknowledgements}

{\it Software used in this work: }

Python \citep{van1995python} and its packages NumPy \citep{harris20}, SciPy \citep{SciPy20}, Pandas \citep{pandas20}, Matplotlib \citep{hunter07}, ColorCET \citep{colorcet}. pyIGIMF is available on PyPI\footnote{\url{https://pypi.org/project/pyIGIMF/}} and on GitHub\footnote{\url{https://github.com/egjergo/pyIGIMF}\label{sec:software}}.

\appendix \label{sec:appendix}

\section{How to use the software}\label{sec:appendix-software}
Further detailed instructions are provided on the Github page (see Footnote~\ref{sec:software}). Users can freely edit the code as needed. To reproduce the default parameters used in this paper, first install the software from the terminal using  {\tt pip}:
\begin{lstlisting}[style=terminal]
    pip install igimf
\end{lstlisting}

To calculate the IGIMF of a galaxy over a given time interval, $t_i + \delta t$, the user must provide the average SFR and average metallicity during that interval. Either the metallicity, {\tt `[Z]'} (Eq.~\ref{eq:metallicity}), or as the metal mass fraction, {\tt `Z'} (Eq.~\ref{eq:metalmassfrac}), is accepted. The user must also provide the stellar mass vector, {\tt mass\_star}. Additionally, the function {\tt igimf.interp} accepts two values for the {\tt alpha1slope} keyword: {\tt `linear'} (Eq.~\ref{eq:alpha1}) and {\tt `logistic'} (Eq.~\ref{eq:alpha1_alternative}). The code returns near-instanteneously the IGIMF vector, as shown in Fig.~\ref{fig:IGIMFsbyZ_linear} and Fig.~\ref{fig:IGIMFsbyZ_logistic}. Below is an example of a possible python script:
\begin{figure}[H]
\begin{lstlisting}[style=terminal, language=Python]
import numpy as np
import pandas as pd
from igimf import classes as inst

Z_solar = 0.0142

# input (edit as needed)
SFR = 2 # Msun/yr
metal_mass_fraction = 0.1 * Z_solar
mass_star = np.logspace(np.log10(0.08), np.log10(150), 100)
alpha1slope = 'logistic' # or 'linear' 

o_IGIMF = inst.IGIMF(metal_mass_fraction=metal_mass_fraction, SFR=SFR, alpha1slope=self.alpha1slope)

igimf_v = o_IGIMF.IGIMF_func(mass_star)
\end{lstlisting}
\end{figure}
\noindent ``{\tt igimf\_v}'' contains the IGIMF associated with a stellar mass vector ``{\tt mass\_star}''. For a stellar mass vector of length 100, the runtime is 0.2 to 0.3 seconds on a standard personal computer.

\section{Variation of the sIMF with Metallicity}

\begin{figure*}[ht]
    \centering \includegraphics[width=\textwidth]{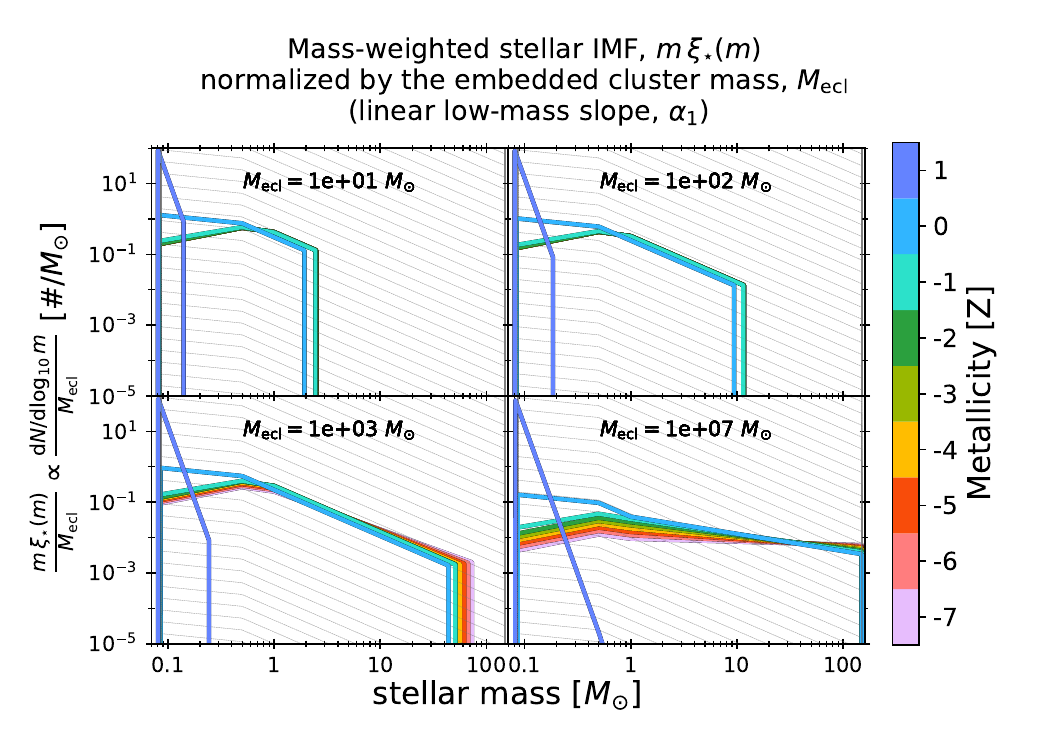}
    \caption{
    Mass-weighted sIMF variations, computed using the linear $\alpha_1(Z)$ slope (Eq.~\ref{eq:alpha1}). For the given metallicity range, these curves are identical to the logistic $\alpha_1(Z)$ slope (Eq.~\ref{eq:alpha1_alternative}). Each panel refers to a different embedded cluster stellar mass, $M_{\rm ecl}$, while the \emph{color-bar} identifies the metallicity, $[Z]$ (Eq.~\ref{eq:metallicity}). The \emph{gray lines} are parallel to the canonical mass-weighted sIMF (Eq.~\ref{eq:canonicalIMF}). Every sIMF is normalized so that the area under each curve equals unity}
    \label{fig:mw_sIMFsbyZ_linear}
\end{figure*}
\begin{figure*}[ht]
    \centering \includegraphics[width=\textwidth]{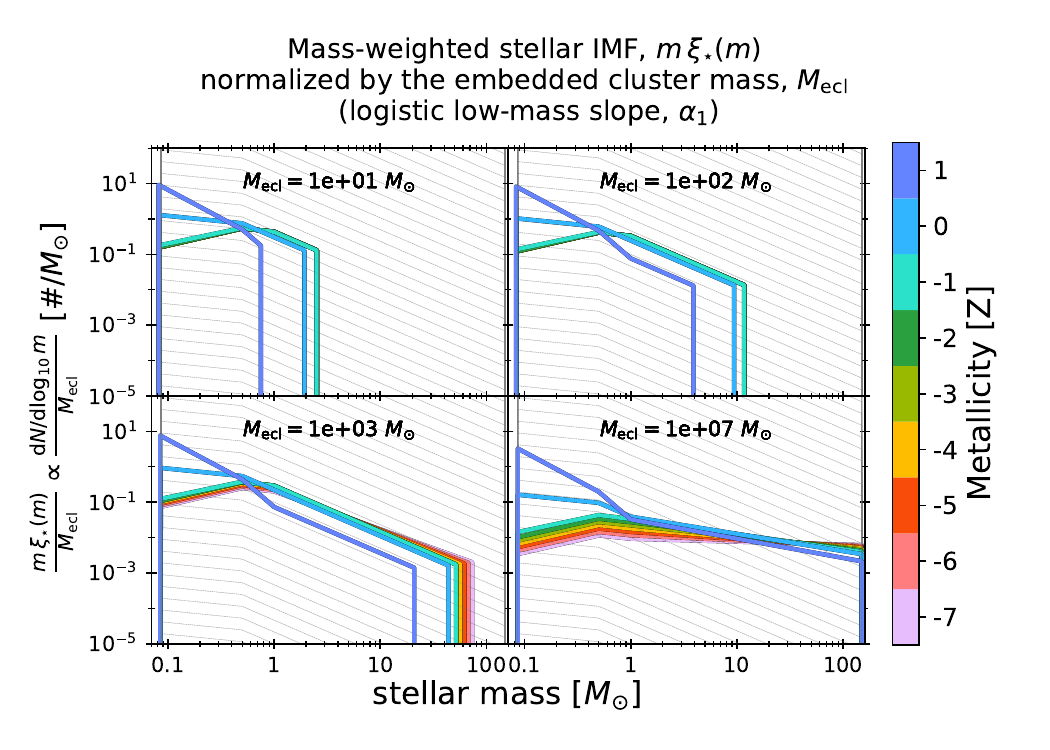}
    \caption{Similarly to Fig.~\ref{fig:mw_sIMFsbyZ_linear}, mass-weighted sIMF variations, computed using the logistic $\alpha_1(Z)$ slope (Eq.~\ref{eq:alpha1_alternative}).}
\label{fig:mw_sIMFsbyZ_logistic}
\end{figure*}

To better compare the dependence of the sIMF with metallicity, we plot  in Fig.~\ref{fig:mw_sIMFsbyZ_linear} and Fig.~\ref{fig:mw_sIMFsbyZ_logistic} for the linear and logistic $\alpha_1(Z)$ prescriptions, respectively. In these plots we follow the convention of using the logarithmic mass as a differential, $\diff \log_{10}(m)$. This is equivalent to tracking a mass-weighted IMF, as explained in Sec.~\ref{sec:mass-weighted}.
\myfig \ref{fig:mw_sIMFsbyZ_logistic} represents the same sIMFs from \myfig \ref{fig:sIMFsbySFR_logistic}, but grouped according to their embedded cluster masses and color-coded according to their metallicity. In this plot it is more clearly evident that an increase in metallicity causes the sIMF to become bottom-heavy. For the lightest-to-intermediate embedded cluster masses, the low gas mass available in the embedded cluster poses a hard limit on the most massive stellar mass generated by that given stellar population.

\begin{figure*}
    \centering
\includegraphics[width=\columnwidth]{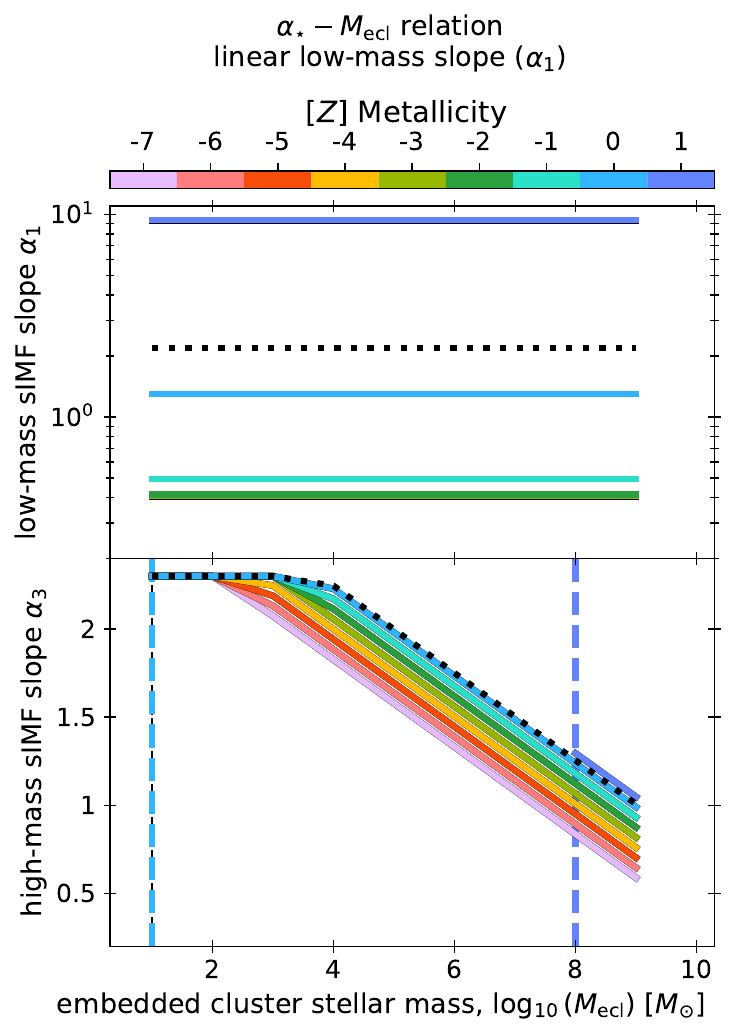}
\includegraphics[width=\columnwidth]{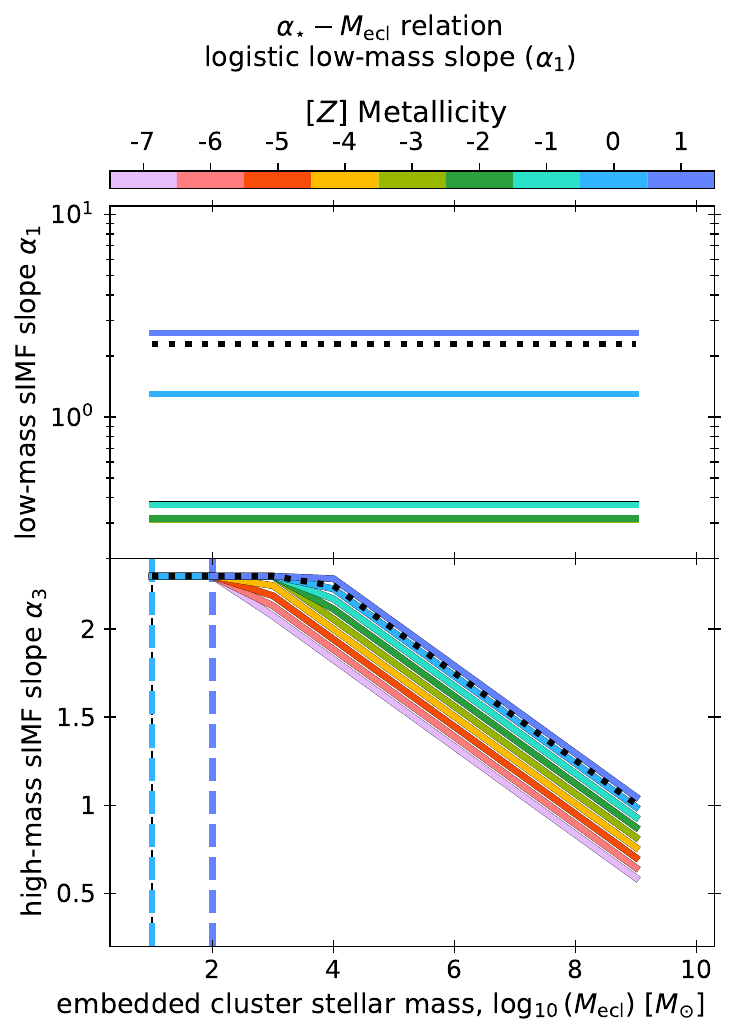}
    \caption{Similar to Fig.~\ref{fig:alpha3alpha1}, but for the full metallicity range.}
    \label{fig:alpha3alpha1_full}
\end{figure*}

In Fig.~\ref{fig:alpha3alpha1_full} we see the $\alpha_1$ and $\alpha_3$ dependence for the full metallicity range, for both the linear and the logistic $\alpha_1(Z)$ prescription. While the $\alpha_1(Z)$ prescription affects the low-mass stars, with increasing metallicity also the high-mass range is affected. We see from the figure that below the black-dotted line at a metal~mass~fraction twice solar, $Z=2 \, Z_{\odot} \approx 0.03$, i.e., $[Z] \approx 0.3$, that both prescriptions manage to produce at least $1 \, M_{\odot}$ stars or above, as shown by the vertical lines. In the case of the logistic prescription, an $M_{\rm ecl} \approx 100 \, M_{\odot}$ is sufficient to produce $1 \, M_{\odot}$ stars even at extreme metallicities of $[Z] = 1$. At the same metallicity, the linear prescription requires $M_{\rm ecl} \approx 10^8 \, M_{\odot}$ to produce $1 \, M_{\odot}$ stars. Below $[Z] = 0.3$, the two prescriptions produce similar $\alpha_1$ slopes. However, the logistic prescription reaches a flatter slope of $\alpha_1 \approx 0.3$ for $[Z] \leq -1$, while the linear prescription reaches $\alpha_1 \approx 0.4$ for the same metallicity.

\section{On Notation conventions}

\subsection{Metallicity}\label{sec:metallicity} Here we define metallicity as the content of all natural elements other than H and He. The metal~mass~fraction is thus defined as:
\begin{equation}\label{eq:metalmassfrac}
    Z = \frac{M_{Z}}{M_{\rm tot}}\, ,
\end{equation}
where $M_Z$ is the total mass in metals, and $M_{\rm tot}$ is the total stellar or gas mass, depending on the context (star, star cluster, or gas cloud). 

Metallicity, $[Z]$, in the standard logarithmic square bracket notation represents the ratio normalized to the solar metal~mass~fraction, $Z_{\odot}$:
\begin{equation}\label{eq:metallicity}
    [Z] = \log_{10}\left(\frac{Z}{Z_{\odot}}\right) \, .
\end{equation}
Using solar metal mass fraction from \citet[][their proto-solar metallicity from their Table 4]{asplund09}, $Z_{\odot} = 0.0142$ the theoretical upper limit of $[Z]$ in the absence of H and He is $[Z] \leq 1.845$ for $Z=1$, i.e., the unrealistic scenario where no H or He are present.

Another common representation of metallicity is expressed relative to hydrogen:
\begin{equation}\label{eq:MvHbymass}
        [Z/H] = \log_{10}\left(\frac{M_Z}{M_H}\right) - \log_{10}\left(\frac{M_Z}{M_H}\right)_{\odot} \, ,
\end{equation}
where the normalization explicitly uses hydrogen mass, $M_H$. In this case, $[Z/H]$ has no theoretical upper limit, and high values of $[Z/H]$ indicate extreme hydrogen depletion. 

Some observational works prefer to define metallicity ratios by number instead of mass, i.e.:
\begin{equation}\label{eq:MvHbynumb}
    [M/H] = \log_{10}\left(\frac{N_M}{N_H}\right) - \log_{10}\left(\frac{N_M}{N_H}\right)_{\odot} \, ,
\end{equation}
where $N_M$ is the sum of all the metal atoms and $N_H$ is the number of hydrogen atoms. 
In the present work, however, we consistently adopt the definition by mass.

\subsection{Mass-weighted IMF and logarithmic mass-scale plots}\label{sec:mass-weighted}
Observational studies sometimes express the sIMF as a logarithmic differential in stellar mass, ${\rm d} N/ {\rm d} \log(m)$ rather than in the linear form $\diff N / \diff m$. The mass-weighted sIMF, $\xi_m(m)$, can be written as:
\begin{equation}
    \xi_m(m) = m \, \xi_*(m) = m \, \frac{{\rm d}N}{{\rm d}m}
\end{equation}
Since $\frac{{\rm d} \log_{10}(x)}{{\rm d}x} = \frac{1}{x \, \ln(10)}$, one obtains:
\begin{align}
    \xi_{\rm log_{10}}(m)&=\frac{{\rm d} N}{ {\rm d} \log_{10}(m)}=\frac{{\rm d}N}{{\rm d} m}\frac{{\rm d} m}{{\rm d} \log_{10}(m)}\\
    &= \frac{{\rm d}N}{{\rm d} m} \, \ln(10) \, m =\ln(10) \, m \, \xi_*(m)
\end{align}
This logarithmic form of the sIMF thus represents the distribution of newly formed stars by mass rather than by number.

\bibliography{IGIMFbib}{}
\bibliographystyle{aasjournal}

\end{document}